\documentclass[preprint]{elsarticle}
\usepackage{fullpage}
\usepackage{graphicx}
\usepackage{booktabs}

\newcommand{\BibTeX}{{\rm B\kern-.05em{\sc i\kern-.025em b}\kern-.08em\ignorespaces
  T\kern-.1667em\lower.7ex\hbox{E}\kern-.125emX}}

\usepackage{amsmath,amssymb}
\journal{ }
\begin{document}
\overfullrule=9pt
%
\begin{frontmatter}
\title{Turing Instability of a Closed Nutrient-Phytoplankton-Zooplankton-Detritus Model\tnoteref{*}}%
\tnotetext[*]{SAC and FJP acknowledge the support of the Natural Sciences and Engineering Research Council of Canada.}
\author[waterloo]{Xiangye Xu} 
\ead{xiangye.xu@uwaterloo.ca}
\author[waterloo]{Sue Ann Campbell}
\ead{sacampbell@uwaterloo.ca}
\author[waterloo]{Francis J. Poulin\fnref{cor}}
\ead{fpoulin@uwaterloo.ca}
\fntext[cor]{Corresponding author.}

%

\affiliation[waterloo]{organization={Department of Applied Mathematics, University of Waterloo},
            city={\\Waterloo},
            state={ON},
            postcode={N2L 3G1}, 
            country={Canada}}




%
%
\begin{abstract}
We study Turing instability in a closed Nutrient-Phytoplankton-Zooplankton-Detritus \newline (NPZD) ecosystem model with nutrient recycling, formulated as a reaction–diffusion system without an explicit time-delay. The detritus compartment represents the biomass during the recycling process and acts to introduce a time delay in this process. Total biomass in the system is conserved, leading to a closed model. 

Although spatial diffusion typically promotes stability, our focus is on understanding how different diffusion rates for different species can destabilize steady states and give rise to spatial patterns. To address this, we first perform linear stability analysis to identify conditions under which Turing instability may occur. We then validate these theoretical predictions through numerical simulations. Our approach develops  both intuition and rigorous analysis of how diffusion influences instabilities in the context of a conservative system, offering a comprehensive framework for understanding pattern formation in multi-species plankton ecosystems.

\vskip .1in
\noindent
{\bf Resum\'e} 
\vskip .1in
\noindent
Nous étudions l'instabilité de Turing dans un modèle écologique fermé
Nutriments, Phytoplancton, Zooplancton et Détritus (NPZD) avec recyclage
des nutriments, formulé sous la forme d'un système de réaction--diffusion
sans retard temporel explicite. Le compartiment des détritus représente
la biomasse au cours du processus de recyclage et introduit ainsi
effectivement un retard dans ce processus. La biomasse totale du système
est conservée, ce qui conduit à un modèle fermé.

Bien que la diffusion spatiale favorise généralement la stabilité,
notre objectif est de comprendre comment des taux de diffusion distincts
pour les différentes espèces peuvent déstabiliser des états stationnaires
et conduire à l'apparition de structures spatiales. À cette fin, nous
effectuons d'abord une analyse de stabilité linéaire afin d'identifier
les conditions dans lesquelles une instabilité de Turing peut se produire.
Nous validons ensuite ces prédictions théoriques à l'aide de simulations
numériques.

Notre approche développe à la fois l'intuition physique et l'analyse
mathématique rigour-euse du rôle de la diffusion dans les mécanismes
d'instabilité au sein d'un système conservatif. Elle fournit ainsi un
cadre complet pour la compréhension de la formation de motifs spatiaux
dans les écosystèmes planctoniques multi-espèces.
\end{abstract}

\end{frontmatter}
\section*{Introduction}
\label{sec:intro}


Planktonic ecosystems form the foundation of marine food webs and play a vital role in the regulation of the global carbon cycle. Due to their wide distribution and strong reproductive capacity, phytoplankton serve as a primary food source for larger organisms, including zooplankton, small fish and other marine species. Consequently, understanding the dynamics of plankton populations is essential to elucidate broader ecological and bio-geochemical processes. To capture the spatial and temporal behavior of planktonic populations more realistically, diffusion terms are included in our mathematical model. These diffusion components represent the spatial dispersal and mixing of plankton concentrations due to the turbulence that occurs in the ocean throughout but especially near the surface \cite{svansson1996diffusion}. This is really trying to capture the subgrid scale processes that are not resolved but need to be parameterized to include turbulent processes.  There is a lot of uncertainty on how to pick these diffusion coefficients but the mechanisms are essential to the dynamics. We show in the context of a planktonic ecosystem, that the inclusion of diffusion can lead to Turing instabilities, and the formation of Turing patterns.  This type of instability is widely observed in nature, for example, in the structures of the bull's-eye colony of Proteus mirabilis \cite{armbruster2012merging}.

Phytoplankton ecosystems play a fundamental role in regulating marine biodiversity and many models describe them using a dynamical systems approach. A list of some earlier approaches can be found in \cite{gentleman2002chronology}.  From 1980 and 1990, as computational power and conceptual frameworks matured, such models gained traction as tools for hypothesis testing. For example, \cite{franks1986prediction} used a closed, three compartmental Nutrient-Phytoplankton-Zooplankton (NPZ) model to study whether the physical decay of warm-core rings could lead to the enhanced phytoplankton growth where the physics and biological dynamics are coupled. Other usage of NPZ models can be found in \cite{franks2002npz}. Furthermore, other closed ecosystem models can be found in \cite{kmet1996material}.

In modeling planktonic dynamics, both closed lakes and marine systems can be considered. A key step in the evolution of these models was the recognition that nutrient cycling is a fundamental process governing system behavior, as previously studied in general ecosystem models \cite{nisbet1976model,ulanowicz1972mass}. Beretta et al.~\cite{beretta1990stability} emphasized that nutrient recycling must be explicitly incorporated in planktonic models, even in the context of chemostat-type formulations. In small-scale ecosystems, the inflow and outflow of biomass (the so-called washout rate) can be relatively large and thus must be accounted for. However, in larger and more self-contained ecosystems, the washout rate becomes negligible, making nutrient recycling within the system the dominant factor. Building on this, other authors analyzed chemostat-type models for Nutrient-Phytoplankton systems, highlighting how delayed recycling affects stability \cite{ruan1995effect,he1998global} and may induces oscillations \cite{ruan2001oscillations}. More recently, \cite{kloosterman2014closed} showed that introducing a weak gamma distributed delay in nutrient recycling within a closed NPZ framework is mathematically equivalent to extending the system to an NPZD model by explicitly including detritus as a fourth compartment. In this formulation, the detritus represents the dead biomass of phytoplankton and zooplankton, which is gradually transformed back into dissolved nutrients, thereby completing the cycle and enriching the ecological realism of the model. Our research is based on this prior work where  planktonic modeling with delayed nutrient recycling is considered in a closed ecosystem.


In systems involving the spatial assembly of particles such as bacteria, chemicals, or biological species, diffusion plays a crucial role in describing how these entities spread over space. Under certain conditions, diffusion-driven instability, also known as Turing instability, may occur \cite{satnoianu2000turing,murraymathematical2}. This phenomenon arises when a spatially uniform steady state, stable in the absence of diffusion, becomes unstable once diffusion is introduced, leading to the emergence of spatially heterogeneous patterns.

Often, diffusion tends to stabilize a system as it smooths the solution based on the concavity at a given instant and spatial location. For example, it is known in the classical diffusion equation that the small spatial scales diffuse more rapidly compared to the larger scales.  Reaction–diffusion systems describe the combined effects of species reactions and diffusion on the dynamics of a system. Interestingly, in such systems, diffusion can paradoxically induce instability, giving rise to organized spatial structures.

Including diffusion in planktonic ecosystem models is essential to accurately describe the evolution of a system, as this describes the effects of turbulence.  Empirical estimates of vertical eddy diffusivity have been obtained from observed phytoplankton and nutrient profiles, allowing parameterization of vertical mixing in models \cite{parker1991eddy}. Recent studies have demonstrated that diffusion can play a more active and nuanced role in planktonic ecosystems than simple homogenization. In particular, \cite{mondal2023impact} examined the impact of eddy diffusivity on seasonal phytoplankton bloom dynamics using a vertically resolved NPZ model. These findings illustrate that diffusion, rather than merely stabilizing spatial distributions, can fundamentally alter system stability and drive transitions in ecosystem behavior.  These studies highlight how field and experimental observations can constrain the diffusion coefficients of nutrients, phytoplankton, and zooplankton, ensuring biologically consistent model parameterizations.

Murray~\cite{murray2007mathematical} provides a general theoretical framework for Turing instability in reaction-diffusion systems of two components, deriving general conditions under which interactions between species lead to instability. Turing instability has been studied in plankton models for several decades, further extending these theoretical insights to marine ecosystems.  For example, the effect of time delay on the Turing instability was studied in a chemostat-type NP plankton model with delayed nutrient recycling \cite{ruan1998turing}.  More recently, \cite{du2010nonlocal} studied reaction–diffusion–advection equations for a single planktonic species by examining the asymptotic profiles of positive steady states under small and large diffusion limits. In addition, \cite{lu2021turing} investigated Turing instability in a two-compartment PZ planktonic model.   Diffusive NPZ models with nutrient recycling have been studied without \cite{giricheva2019spatiotemporal} or with time delay~\cite{tao2021dynamics}. 

Recent work has explored Turing-type mechanisms in more realistic open or partially coupled systems, where boundary fluxes and external forcing play a crucial role in pattern selection~\cite{krause2021isolating}. Extensions to reaction diffusion models further reveal that coupling between diffusive and non-diffusive species can induce Turing patterns~\cite{marciniak2017instability}. Moreover, studies in predator–prey systems with nonlinear cross-diffusion and Allee effects demonstrate that Turing instability can emerge under biologically plausible parameter regimes \cite{shao2022turing}, while investigations in spatially heterogeneous domains show that environmental variability strongly influences pattern onset and morphology~\cite{van2021pattern}. Moreover, \cite{villar2023general} established the general conditions under which an $n$-component reaction diffusion system can exhibit Turing instability, and \cite{villar2025designing} extended this work by providing a constructive framework for designing $n$-component systems that produce such instabilities through appropriate choices of diffusion tensors and linearized reaction kinetics.

Motivated by our previous work
\cite{kloosterman2014closed,tao2021dynamics}, we consider a diffusive model with four components: nutrient, phytoplankton, zooplankton and detritus.  Nutrient  recycling is incorporated into the model in such a way that 
the total biomass in the system is conserved. 
This, together with diffusion, 
provide what seem to be stabilizing forces in the model. Nevertheless, we show that having different diffusion scales in the different components can potentially cause Turing instability. To our knowledge, no previous work has studied Turing instability in a four compartment (NPZD) plankton model. 

The outline of our article is as follows. In section 2 we define the NPZD model that we consider throughout and introduce the conservation laws of total biomass. In section 3, we apply linear stability analysis separately to the ODE model 
without diffusion and the PDE model 
with diffusion.  This yields a  condition through which Turing instability can occur in the NPZD model. In section 4, we numerically simulate the plankton model to validate the qualitative conditions and explore the nonlinear dynamics.  Finally, in section 5 we summarize and discuss our findings.

\section{Mathematical Model}
\label{sec:analysis}

Planktonic ecosystem dynamics have been modelled using a variety of compartmental frameworks, one of the most widely studied being the NPZ model introduced by Franks et al.~\cite{franks1986prediction}. In its original formulation, the model consists of three coupled nonlinear ordinary differential equations describing the interactions among nutrients, phytoplankton, and zooplankton. A limitation of many such models is that biomass transfer between compartments is assumed to occur instantaneously. In reality, ecological processes such as mortality, decomposition, and nutrient recycling require finite amounts of time, introducing delays into the system dynamics. Incorporating explicit time delays generally leads to more realistic models, but also significantly increases their mathematical complexity.

Kloosterman et al.~\cite{kloosterman2014closed} demonstrated that, within a closed NPZ framework, the introduction of a distributed time delay with a particular kernel can be reformulated exactly as an NPZD model that includes an explicit detritus compartment. In this interpretation, the detritus serves as an intermediate reservoir that stores biomass prior to its return to the nutrient pool, thereby representing the delayed recycling process. This equivalence provides additional justification for the NPZD formulation, as it captures the effects of delayed nutrient recycling while remaining within the framework of ordinary differential equations rather than delay differential equations.

We consider the NPZD model introduced in \cite{kloosterman2014closed}, augmented with spatial diffusion,
\begin{equation}
\label{eq:NPZD}
\begin{aligned}
\frac{\partial N}{\partial t} &= -\nu P\frac{N}{N+1} + \alpha D + d_1 \frac{\partial^2 N}{\partial x^2}, \\
\frac{\partial P}{\partial t} &= \nu P\frac{N}{N+1} -  gZ\frac{P}{P+1}  - \lambda P + d_2 \frac{\partial^2 P}{\partial x^2},
\\
\frac{\partial Z}{\partial t} &=  gZ\frac{P}{P+1} -\delta Z + d_3 \frac{\partial^2 Z}{\partial x^2},
\\
\frac{\partial D}{\partial t}
&=
\lambda P
+\delta Z
-\alpha D
+d_4 \frac{\partial^2 D}{\partial x^2},
\end{aligned}
\end{equation}
where we assume Michaelis-Menten, also called Holling Type II \cite{gentleman2002chronology} functional forms.  Also, as in \cite{kloosterman2014closed,poulin2010size}, we assume half-saturation constants in terms of uptake and grazing to have values of one.

In the model equations, $N(x,t)$, $P(x,t)$, $Z(x,t)$, and $D(x,t)$ denote the concentrations of nutrients, phytoplankton, zooplankton, and detritus, respectively. These quantities vary in both time $t$ and space $x$. For simplicity, we restrict our attention to a one-dimensional spatial domain representing the vertical structure of the water column. Although this setting captures the essential mechanisms responsible for the instability, extensions to two- and three-dimensional domains remain an important direction for future study.

Each component is associated with its own diffusion coefficient, denoted by $d_1$, $d_2$, $d_3$, and $d_4$. In contrast to the assumptions made by Tao et al.~\cite{tao2021dynamics}, where all diffusion coefficients were taken to be equal and diffusion was shown to have only a stabilizing effect, we allow the diffusivities to differ between components. As we demonstrate below, this distinction is crucial, since differential diffusion can destabilize an otherwise stable equilibrium and lead to the emergence of Turing patterns.

Each compartment in the model represents a distinct reservoir of biomass within the ecosystem. Biomass enters a compartment through growth or reproduction processes and leaves through mortality, predation, or other loss mechanisms. Because the model represents a closed and conservative system, any biomass removed from one compartment is transferred to another, ensuring that the total biomass remains constant over time.

The model includes a nutrient uptake term that transfers biomass from the nutrient compartment to the phytoplankton compartment through photosynthetic growth. While this process depends on the availability of light in natural systems, light limitation is not included in the present simplified formulation. The model also incorporates grazing, whereby phytoplankton biomass is consumed and converted into zooplankton biomass. Although conversion inefficiencies could be represented by routing a fraction of the grazed biomass directly to the detritus pool, such effects are neglected here. In addition, natural mortality of both phytoplankton and zooplankton transfers biomass to the detritus compartment, where it is subsequently recycled back into the nutrient pool, thereby completing the nutrient cycle.

\subsection{Initial and Boundary Conditions}

We impose Neumann boundary condition for each species, which ensures that no biomass is lost through the boundary:
\begin{align}
    \frac{\partial}{\partial t}[N, P, Z, D](x, t) = [0, 0, 0, 0]  \quad \mbox{for} \quad x=0,2\pi, \quad \mbox{ and } \quad t\ge 0. \label{eq:BC}
\end{align}
Our initial conditions are as follows,
\begin{align}
    [N, P, Z, D](x, 0) = [N_0, P_0, Z_0, D_0](x), \quad\mbox{for } x\in [0,2\pi]. \label{eq:IC}
\end{align}
We will explore some specific forms for the functions $N_0(x),P_0(x),Z_0(x),D_0(x)$ in section~\ref{sec:results}. 

\subsection{Conservation Laws}

If diffusion is ignored, then equations \eqref{eq:NPZD} reduce to a system of ordinary differential equations, similar to the model introduced in \cite{kloosterman2014closed},
\begin{equation}\label{eq:NPZD_ODE_system}
\begin{aligned}
        \frac{dN}{d t}&= -\nu P\frac{N}{N+1}+ \alpha D,\\
        \frac{dP}{d t}&= \nu P\frac{N}{N+1}-\lambda P - gZ\frac{P}{P+1}, \\
        \frac{dZ}{d t}&= gZ\frac{P}{P+1}-\delta Z, \\
        \frac{dD}{d t}&=  \lambda P + \delta Z -\alpha D.
\end{aligned}
\end{equation}
Adding these differential equations, we obtain the conservation law:
\begin{align}
   \frac{d}{d t}(N + P + Z + D) =0  \label{eq:ODEconslaw}
\end{align} hence, the total biomass is conserved and we define it as
\begin{align}
    N_T = N(t)+P(t)+Z(t)+D(t).  \label{eq:TotalBiomass}
\end{align}
The total biomass will be used in section~\ref{sec:results} as a parameter we vary to observe different behaviors in the system. It is critical to determining the linear stability of the equilibria, as we will see in the next subsection. 

Note that equation~\eqref{eq:TotalBiomass} defines an invariant hyperplane for the system \eqref{eq:NPZD_ODE_system}. 
Given an initial condition, the value of total biomass $N_T$ will be fixed and the solution will evolve on this hyperplane for $t\ge 0$. The dynamics on this hyperplane can be determined by expressing $D$ in terms of the other variables and hence replacing \eqref{eq:NPZD_ODE_system} by the reduced system 
\begin{equation}\label{eq:NPZD_ODE_system3D}
\begin{aligned}
         \frac{d N}{d t}&= -\nu P\frac{N}{N+1}+ \alpha (N_T-N-P-Z),\\
        \frac{d P}{d t}&= \nu P\frac{N}{N+1}-\lambda P -gZ\frac{P}{P+1}, \\
        \frac{d Z}{d t}&= gZ\frac{P}{P+1}-\delta Z,\\
         D(t)&=N_T-N(t)-P(t)-Z(t).
\end{aligned}
\end{equation}

When we sum the PDE model \eqref{eq:NPZD}, we obtain a different equation that is similar to a diffusion equation but it is different because we have a weighted sum of the terms on the right hand side, which is not precisely what appears on the left hand side,
\begin{align*}
       \frac{\partial}{\partial t}[N+P+Z+D] &=  \frac{\partial^2 }{\partial x^2}[d_1 N +d_2 P + d_3 Z + d_4 D]. 
\end{align*}
If we integrate this equation over the spatial domain and impose the boundary conditions from equation \eqref{eq:BC} we determine that the total amount of biomass is conserved in time, and hence the system is said to be closed,
\begin{align*}       
       \frac{\partial}{\partial t} \int_{0}^{2\pi} [N+P+Z+D]\,dx = 0.
\end{align*}
We integrate this equation with respect to time and introduce a new constant for the total biomass at each time,
\begin{equation} 
 \int_{0}^{2\pi}\left[N+P+Z+D\right]\,dx =\tilde{N}_T.
\label{eq:PDEconslaw}
\end{equation}

\section{Linear Stability Analysis}\label{sec:stability}

In this section, we investigate the stability of the equilibrium solutions of both the ODE system \eqref{eq:NPZD_ODE_system}, obtained by neglecting diffusion, and the corresponding spatially homogeneous equilibrium solutions of the reaction diffusion system \eqref{eq:NPZD}. Although the model contains many parameters whose influence could be explored, our primary focus is on the diffusion coefficients. We examine how differential diffusion among the system components can destabilize an otherwise stable homogeneous equilibrium and thereby induce Turing instabilities in the NPZD model. This analysis allows us to identify the conditions, valid for any values of the reaction model parameters, under which diffusion acts not as a stabilizing mechanism, but rather as a driver of spatial pattern formation.

\subsection{Stability without diffusion} \label{subsec:ODEstab}

Let $N^*,P^*,Z^*,D^*$ denote the components of an equilibrium solution of the ODE model, equation \eqref{eq:NPZD_ODE_system}, which are defined by the following system of algebraic equations,
\begin{align}
-\nu P^*\frac{N^*}{N^*+1}+ \alpha D^*&=0\label{eq:eqmdef1},\\
P\left( \nu\frac{N^*}{N^*+1}-\lambda  -gZ^*\frac{1}{P^*+1}\right) &=0, \label{eq:eqmdef2}\\
Z^*\left( g\frac{P^*}{P^*+1}-\delta \right)&=0\label{eq:eqmdef3},\\
 N^*+P^*+Z^*+D^* &= N_T.\label{eq:eqmdef4}
\end{align} 

Any equilibrium solution of the ODE system is also a spatially homogeneous equilibrium solution of the PDE system, since the diffusion terms vanish in the absence of spatial gradients. The only difference is that the conservation law \eqref{eq:eqmdef4} for the ODE system is replaced for the PDE system by
\[ \tilde{N}_T=\int_0^{2\pi} \left[N^*+P^*+Z^*+D^*\right] \,dx = 2\pi N_T.\]
Since the parameter $\tilde{N}_T$ is a multiple of $N_T$, we will only refer to $N_T$ when dealing with equilibria, whether for the PDE system or ODE system. 

We begin by summarizing several results concerning the existence and stability of equilibrium solutions of the model, which follow from the analysis of Kloosterman et al.~\cite{kloosterman2014closed}. Several types of equilibrium solutions can arise depending on the parameter values, but our primary interest lies in those for which both phytoplankton and zooplankton persist. These coexistence states are the most biologically relevant and provide the foundation for the pattern-forming instabilities studied later in this work.  We will call an equilibrium solution {\em biologically feasible} if all the components are nonnegative.

The simplest equilibrium solution, referred to as the \emph{trivial equilibrium}, is
\[
E_0=(N_T,0,0,0),
\]
for which all biomass resides in the nutrient compartment.  Clearly this equilibrium is biologically feasible if $N_T>0$. 

A second class of equilibrium solution corresponds to the absence of zooplankton, i.e., where nutrients, phytoplankton, and detritus coexist. This equilibrium solution is given by
\[
E_{NPD}=(N^*,P^*,0,D^*),
\]
where
\begin{equation}
N^*=\frac{\lambda}{\nu-\lambda}, \quad
P^*=\frac{N_T-\frac{\lambda}{\nu-\lambda}}{1+\frac{\lambda}{\alpha}},\quad
D^*=\frac{N_T-\frac{\lambda}{\nu-\lambda}}{1+\frac{\alpha}{\lambda}}.
\label{NPeqmdef} \end{equation}
This equilibrium is biologically feasible if
\[
\nu>\lambda\quad \mbox{and} \quad \quad N_T>\frac{\lambda}{\nu-\lambda}\stackrel{def}{=}N_{T1}.\label{eq:NPDcond}
\]
Further, $N^*$ is independent of $N_T$ while $P^*,D^*$ are linear, increasing functions of $N_T$. 

With both phytoplankton and zooplankton present, $P^*, Z^* \neq 0$, we have the third type of equilibrium solution $E_{NPZD}=(N^*,P^*,Z^*,D^*)$, which is the focus of our research. The following equations determine this third type of equilibrium:
\begin{align}
    &P^* = \frac{\delta}{g-\delta}\label{eq:eqmP}, \\ 
    & Z^*=\frac{1}{g-\delta} \left(\nu\frac{N^*}{N^*+1}-\lambda\right) \label{eq:eqmZ}, \\
    &D^*=\frac{\nu}{\alpha} \frac{\delta}{g-\delta} \frac{N^*}{N^*+1} \label{eq:eqmD}, \\
    & N^*+P^*+Z^*+D^* = N_T. \label{eq:eqmN}
\end{align}
Straightforward analysis shows that there is a unique biologically feasible equilibrium of this form, if the following conditions are satisfied:
\begin{equation} g>\delta,\quad N_T>N_{T2}=\frac{\lambda}{\nu-\lambda}+\left(1+\frac{\lambda}{\alpha}\right)\frac{\delta}{g-\delta}. \label{eq:NPZDcond}\end{equation}
If either of these conditions is not satisfied, the equilibrium is not biologically feasible. In addition, $P^*$ is independent of $N_T$ while $N^*,Z^*,D^*$ are linear, increasing functions of $N_T$. See \ref{app:eqm} for the details.  
The first condition in \eqref{eq:NPZDcond} corresponds to the maximum growth rate of the zooplankton being larger than the maximum death rate, while the second condition corresponds to the total biomass being sufficiently large.

Now we consider the linear stability of these equilibrium solutions. It is straight forward to check that the Jacobian matrix of the ODE system  eq.~\eqref{eq:NPZD_ODE_system} always has a zero eigenvalue because the system is conservative, which is to say that the total biomass in the system is conserved with respect to time. The zero eigenvalue relates to the property that no movement can occur in the phase space transverse to the invariant hyperplane \eqref{eq:TotalBiomass}. The stability of an equilibrium solution within the hyperplane is determined by the other three eigenvalues. Thus, if these eigenvalues all have negative real parts, then the equilibrium will be stable in the nonlinear system \eqref{eq:NPZD_ODE_system} and asymptotically stable in the reduced system \eqref{eq:NPZD_ODE_system3D}. 

The stability of the $E_0$ and $E_{NPD}$ equilibria is studied in Appendix~\ref{app:eqm}. Using those results and noting the dependence of the equilibria on the total biomass, $N_T$, we have the following summary:
\begin{itemize}
\item $0<N_T<N_{T1}$: $E_0$ is the only biologically feasible equilibrium and it is stable.
\item $N_{T1}<N_T<N_{T2}$: There are two biologically feasible equilibria: $E_0$ and $E_{NPD}$. The former is unstable and the latter is stable. The two equilibria coincide when $N_T=N_{T1}$.
\item $N_{T2}<N_T$: There are three biologically feasible equilibria: $E_0$, $E_{NPD}$ and $E_{NPZD}$. The first two are unstable. The stability of the third is considered below. The $E_{NPZ}$ and $E_{NPZD}$ coincide when $N_T=N_{T2}$.
\end{itemize}

The Jacobian of the ODE system \eqref{eq:NPZD_ODE_system}, evaluated at the equilibrium $[N^*, P^*, Z^*, D^*]$ and simplified using eq.~\eqref{eq:eqmdef4}, is given by:
\begin{equation}
    \label{eq:npzd_jacobian}
    J_{ode}  
    = \begin{bmatrix}
        -\nu P^* \frac{1}{(1+N^*)^2} & -\nu \frac{N^*}{N^*+1} & 0 & \alpha \\
        \nu P^* \frac{1}{(1+N^*)^2} & \nu \frac{N^*}{N^*+1} - \lambda  -g Z^* \frac{1}{(1+P^*)^2} & -\delta & 0 \\
        0 & g Z^* \frac{1}{(1+P^*)^2} & 0 & 0 \\
        0 & \lambda & \delta & -\alpha
    \end{bmatrix},
\end{equation}
with the corresponding characteristic polynomial:
\begin{equation*}
       p_{ode}(\lambda) =\lambda^4 + a_3 \lambda^3 + a_2 \lambda^2 + a_1 \lambda +a_0.
\end{equation*}

Let $J^*_{ode}$ denote $J_{ode}$ evaluated at the equilibrium. Since the model is conservative, the Jacobian has some special structure. We highlight this by writing it in the following simplified form:
\begin{equation}
    J_{ode}^*  = \begin{bmatrix}
        -A & -B & 0 & E\\
        A & B-C-F & -G & 0\\
        0 & F & 0 & 0 \\
        0 & C & G & -E
    \end{bmatrix},
\end{equation}
where we define
\begin{align}
    A =  \nu P^* & \frac{1}{(N^*+1)^2},\ 
    B = \nu \frac{N^*}{N^*+1},\  
    C = \lambda,  \\
    F &= g Z^* \frac{1}{(P^*+1)^2},\  
    E = \alpha,\ 
    G = \delta=g. 
\label{eq:ABdef}\end{align}
Note that these are all positive constants.
The coefficients of the characteristic polynomial then become:
\begin{align*}
    a_0 &=  ADEG-ADEG=0,\\
    a_1 &= AFE+AFG+FEG, \\
    a_2 &= AC+AF+AE - BE +CE +FE+FG, \\
    a_3 &= A-B+C+F+E.
\end{align*} 

The vanishing of $a_0$, which is the determinant of the Jacobian, is a consequence of the fact the model has a conservation law. This means that the characteristic polynomial simplifies as follows: 
\begin{equation}
       p_{ode}(\lambda) =\lambda(\lambda^3 + a_3 \lambda^2 + a_2 \lambda + a_1),
\label{eq:ODEchareq}
\end{equation}
and always has a zero root as discussed above. 

For the equilibrium to be linearly stable, the other roots of \eqref{eq:ODEchareq} must have negative real part. 
To determine conditions for this to occur, we apply the  Routh–Hurwitz stability criterion to
the cubic factor $\lambda^3 + a_3 \lambda^2 + a_2 \lambda + a_1$ \cite{ogata2010modern}. This give rise to the following conditions:
\begin{equation}
\begin{aligned}
    a_1 &> 0,  \\
    a_3 &> 0,  \\
    a_2 a_3 -a_1& >0,
\end{aligned} \label{eq:RH_ODE}
\end{equation}
which guarantee that the equilibrium of the system \eqref{eq:NPZD_ODE_system} is stable. 
Note that the conditions \eqref{eq:RH_ODE} guarantee that the equilibrium is asymptotically stable in the reduced ODE system \eqref{eq:NPZD_ODE_system3D}.

A complete analysis of how the signs of the coefficients depend on the model parameters is not the within the scope of this work, however, we note that $a_1>0$ and the condition $A-B+C+F>0$ will guarantee that both $a_2>0$ and $a_3>0$.


\subsection{Diffusion induced instability} \label{subsec:PDEstab}
Now we return to the model with diffusion~\eqref{eq:NPZD}. We consider spatially homogeneous equilibria, i.e., solutions that satisfy 
\[ [N(x,t),P(x,t),Z(x,t),D(x,t)]=[N^*,P^*,Z^*,D^*], \] 
for all time and $x$ in our spatial domain of $[0, 2\pi]$. It is easy to check that $N^*,P^*,Z^*,D^*$ are determined by the same equations as for equilibria of the ODE system, i.e, equations~\eqref{eq:eqmP}-\eqref{eq:eqmN}. 

The linearization of  \eqref{eq:NPZD} around the equilibrium is
\begin{equation}
         \frac{\partial W}{ \partial t}  = J^*_{ode} W+ {\mathcal D} \nabla W,
\label{eq:linear_approximation}
\end{equation} 
where 
\begin{align*}
    {\mathcal D}&=\begin{bmatrix}
        d_1 & 0 & 0 & 0 \\
        0 & d_2 &0 & 0\\
        0 & 0 & d_3 & 0\\
        0 & 0 &0 & d_4
\end{bmatrix}.
\end{align*}

Based on the boundary condition \eqref{eq:BC}, we assume a solution in the form:
\begin{equation}
    W = \sum_{k\in {\mathcal K}} \exp(\lambda_k t) \cos(k x)\, {\bf v}_k,
    \label{eq:perturbation_assumption}
\end{equation} 
where ${\mathcal K}=\{ \frac{n}{2}, n\in Z^+\}$. The values in $\mathcal K$ follow from the Neumann boundary conditions, equations \eqref{eq:BC}, and the fact that our spatial domain is $[0,2\pi].$

Substituting the trial solution into \eqref{eq:linear_approximation} gives the eigenvalue problem 
$$
\lambda_k {\bf v}_k = J^*_{pde} {\bf v}_k,
$$ where the
simplified version of the PDE Jacobian is given by:
\begin{equation} \label{eq:pde_jacobian}
    J_{pde}^*=J_{ode}^*-k^2 {\mathcal D}  = \begin{bmatrix}
        -A-d_1 k^2  & -B & 0 & E\\
        A & B-C-F-d_2 k^2  & -G & 0\\
        0 & F & -d_3 k^2 & 0 \\
        0 & C & G & -E-d_4 k^2 
    \end{bmatrix}.
\end{equation}

The corresponding characteristic polynomial is:
\begin{equation}
       p_{pde}(\lambda) =\lambda^4 + b_3 \lambda^3 + b_2 \lambda^2 + b_1 \lambda +b_0,
\label{eq:PDEchareq}
\end{equation}
with coefficients in the form 
\[ b_j=a_j+h_j(k^2,d_1,d_2,d_3,d_4,A,B,C,E,F,G).\] 

Since we are interested in Turing instability, we focus on how the presence of different diffusion constants may induce instability of the equilibrium, given that it is stable for the ODE system, i.e., that conditions~\eqref{eq:RH_ODE} hold.  From simple continuity arguments, the characteristic polynomial~\eqref{eq:PDEchareq} will have a real, positive root if $b_0<0$. Further since, $a_0=0$ and  $a_j>0,\ j=1,\ldots,3$ from ~\eqref{eq:RH_ODE}, the easiest path to instability is through $b_0$. Thus we focus on this coefficient, which is given by
\begin{align*} 
b_0 =& k^2 (AFE d_3 + AFG d_4 + FEG d_1)\\
    & + k^4 \bigg( AC d_3 d_4 +AF d_3 d_4 +AE d_2 d_3 +(-B+C+F)E d_1 d_3  +FG d_1 d_4  \bigg) \\
    &+ k^6 \bigg( A d_2 d_3 d_4 +(-B+C+F) d_1 d_3 d_4 +E d_1 d_2 d_3 \bigg) \\
    & + k^8 d_1 d_2 d_3 d_4.
\end{align*}
Recall that $A,B,C,E,F,G$ and the diffusion constants $d_j$ are all positive. Thus, the coefficients of $k^2$ and $k^8$ are positive, and $b_0<0$ will only be possible if the coefficients of $k^4$ and/or $k^6$ are negative.

Next, consider the term $-B+C+F$. Recalling the definitions of $B,C,F$ from eq.~\eqref{eq:ABdef},
and using eq.~\eqref{eq:eqmdef2}, 
we have
\begin{align*}
    -B+C+F & = -\nu \frac{N^*}{N^*+1} + \lambda + g Z^* \frac{1}{(P^*+1)^2}, \\
    & = -\lambda -gZ^*\frac{1}{P^*+1}  + \lambda  + g Z^* \frac{1}{(P^*+1)^2}, \\
    & =  -gZ^*\frac{P^*}{(P^*+1)^2}.
\end{align*}
Thus $-B+C+F<0$ {\em for any} values of the model parameters. 

Thus, the coefficients of $k^2$ and $k^4$ may be negative for appropriate choices of the diffusion constants. If we assume the following conditions,
\begin{equation}
  \begin{aligned}
    d_1 &\gg
    d_3 \gg d_2, d_4. 
  \end{aligned}
\label{eq:instability_condition}
\end{equation}
then the coefficients are dominated by the following terms:
\begin{equation}
\begin{aligned}
k^4 &: (-B+C+F)E d_1 d_3 , \\
 k^6 &: (-B+C+F) d_1 d_3 d_4 +E d_1 d_2 d_3. 
 \end{aligned}
 \label{eq:cofficients_b02}
\end{equation}
Since $-B+C+F<0$ the coefficient of $k^4$ will be negative.
We conclude that  instability is possible if the $k^4$ terms dominate in $b_0$, i.e., $k$ is not too small or too large.

In summary, for any values of the ODE model parameters satisfying eq.~\eqref{eq:RH_ODE}, Turing instability is possible if the diffusion constants satisfy \eqref{eq:instability_condition}. In the next section we will use this analysis to find parameter values such that Turing instability occurs and numerically study the influence of the several parameters on this instability. 



\section{Numerical Results}
\label{sec:results}

\subsection{Methods}

Our numerical investigations are carried out using the Julia programming language~\cite{Julia-2017}. For the linear stability analysis, we use the \texttt{LinearAlgebra} standard library to compute the eigenvalues of the Jacobian matrices $J_{\mathrm{ODE}}^*$ and $J_{\mathrm{PDE}}^*$. To verify the theoretical predictions and provide numerical evidence of Turing instability, we perform direct simulations of the reaction--diffusion system using the \texttt{Oceananigans} framework~\cite{OceananigansJOSS}. Although \texttt{Oceananigans} was developed as a high-performance finite-volume solver for nonhydrostatic ocean dynamics, it also provides a flexible platform for solving reaction--diffusion equations. In the present study, we employ the library to solve the one-dimensional NPZD reaction--diffusion system and explore its nonlinear dynamics.

The computational domain is chosen to be $[0,2 \pi]$ is discretized on a rectilinear staggered grid with $64$ points, and this should be thought of as the vertical direction. The grid topology is set as \texttt{(Flat, Flat, Bounded)}, indicating no spatial components in $x$ and $y$ and the $z$-direction is bounded at the top and bottom, where we impose Neumann boundary conditions as in equation \ref{eq:BC}.
Time integration is performed using a third-order Runge--Kutta scheme. To ensure stability according to the Courant--Friedrichs--Lewy (CFL) condition \cite{courant1967partial}, the time step is defined as
\[
\Delta t = \frac{ 0.1  (\Delta z)^2}{d_M},
\]
where $\Delta z$ is the vertical grid spacing size and $d_M$ is the largest diffusivity coefficient in the system.

\subsection{Parameters}

Table~\ref{tab:npzd-params} gives the parameter values used in the numerical simulations. The first four are taken from \cite{kloosterman2014closed,poulin2010size}. The detritus transition rate is chosen to be $\alpha=1$,  which is consistent with the remineralization rate of detritus by bacteria used by various authors \cite{davis1994biological,Edwards2001}.  We acknowledge there is a lot of uncertainty in these parameters.  We are not trying to describe a particular system, but instead learn about the types of motions that can develop and in particular how Turing instabilities might manifest themselves in models of plantonic ecosystems. Since the analysis of the previous section applies for any values of the reaction model parameters, we chose a set of biologically reasonable values from our previous work.  To see a different set of parameter values that could be considered, the reader is directed to \cite{taniguchi2014planktonic}. Given the parameters of Table~\ref{tab:npzd-params}, the values for $N_T,\ d_1,\ d_2,
 d_3$ and $d_4$ are chosen to ensure the possibility of Turing instability and will be discussed in sections~\ref{subsec:PDEstab} and \ref{subsec:PDEstabn}. 

\begin{table}[ht]
\centering
\caption{Model Parameters. Values in the bracket $()$ represent values used to illustrate Turing instability.}
\begin{tabular}{|c|c|c|}
\hline
\textbf{Parameter} & \textbf{Symbol} & \textbf{Value} \\ \hline
{Phytoplankton maximum birth rate} & \( \nu \) & $5.9$  day$^{-1}$\\ \hline
{Phytoplankton maximum death rate} & \( \lambda \) & $0.017$ day$^{-1}$ \\ \hline
{Zooplankton maximum grazing rate} & \( g \) &  $7$   day$^{-1}$\\ \hline 
{Zooplankton maximum death rate} & \( \delta \) & $0.17$  day$^{-1}$\\ \hline\hline
{Detritus transition rate} & \( \alpha \) & \( 1 \)  day$^{-1}$\\ \hline
{Total biomass} & \( N_T \) & varied ($0.8$) \\ \hline
{Nutrient diffusion coefficient } & \( d_1 \) & varied $(2)$ \\ \hline
{Phytoplankton diffusion coefficient} & \( d_2 \) & \( 0.0001 \) \\ \hline
{Zooplankton diffusion coefficient} & \( d_3 \) & \( 0.001 \) \\ \hline
{Detritus diffusion coefficient} & \( d_4 \) & \( 0.0001 \) \\ \hline
\end{tabular}

\label{tab:npzd-params}
\end{table}

\newpage

\subsection{Stability without diffusion}
\begin{figure}[t]
\begin{center}
\includegraphics[height=0.4\textwidth,angle=270]{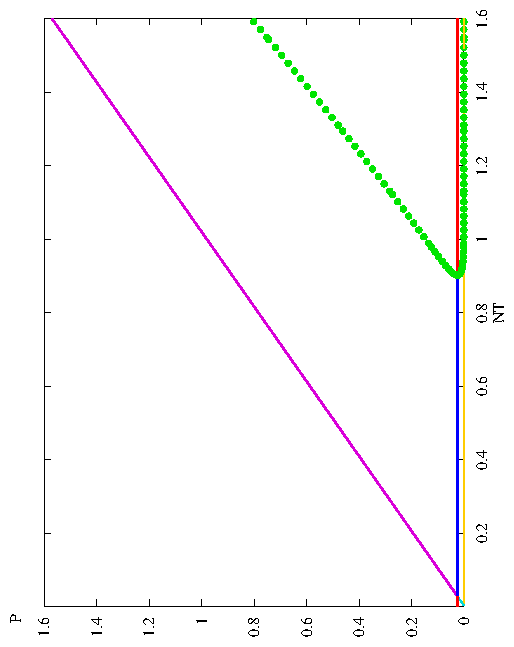}
\includegraphics[height=0.4\textwidth,angle=270]{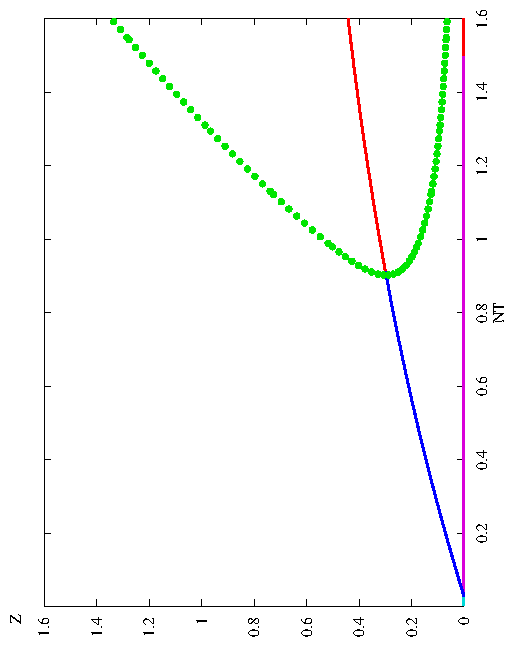}
\caption{Bifurcation Diagrams for the ODE model showing magnitude of $P$ and $Z$ as $N_T$ is varied. Other parameter values are given in Table~\ref{tab:npzd-params}. Blue/red lines correspond to stable/unstable $E_{NPZD}$ equilibria. Green dots correspond to maximum/minimum of stable NPZD periodic orbit. Cyan/magenta lines correspond to stable/unstable $E_{NPD}$ equilibria. Orange line corresponds to unstable $E_{0}$ equilibria.}
\label{fig:ODE_stability}
\end{center}
\end{figure}

Recall from section~\ref{subsec:ODEstab} that the conservation law introduces the total biomass, $N_T$, as a parameter in the equations defining the equilibria of the model. Thus the stability of the system is dependent on the value of this parameter.   To study this, we use the numerical continuation software AUTO \cite{Auto} in the package XPPAUT \cite{XPPAUT} to make a bifurcation diagram of the reduced system \eqref{eq:NPZD_ODE_system3D}. We fixed the model parameters as in Table~\ref{tab:npzd-params} and used $N_T$ as the bifurcation parameter. See Figure~\ref{fig:ODE_stability}. As predicted by the analysis, there are transcritical bifurcations at $N_T=N_{T1}=0.00289$, leading to the emergence of the asymptotically stable $(N^*,P^*,0,D^*)$ equilibrium,  and at $N_T=N_{T2}=0.0282$ leading to the emergence of the asymptotically stable $(N^*,P^*,0,D^*)$ equilibrium. Further at $N_T=0.9018$ the $(N^*,P^*,0,D^*)$ equilibrium loses stability in a supercritical Hopf bifurcation.

\subsection{Diffusion induced instability}\label{subsec:PDEstabn}

\begin{figure}[t]
\begin{center}
   \includegraphics[width=\textwidth]{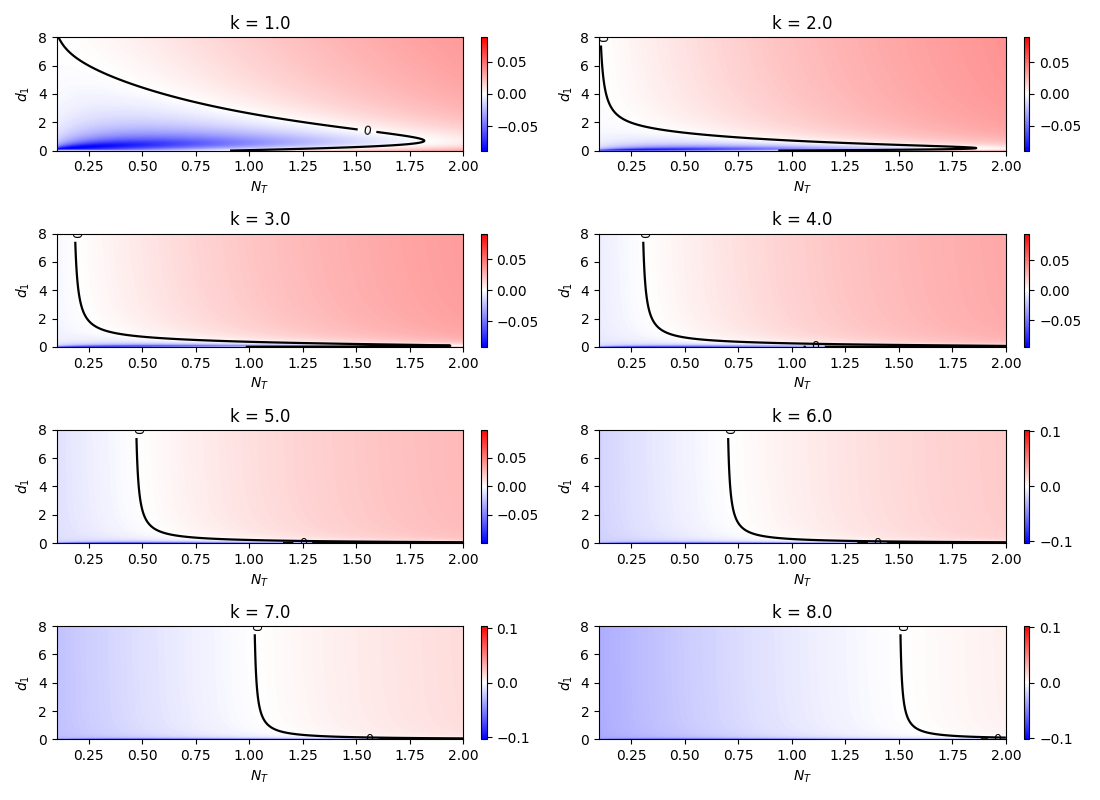}
    \caption{Dependence of linear growth rate of PDE model \eqref{eq:NPZD} on total biomass $N_T$ and phytoplankton diffusion rate $d_1$ for modes $k$, as computed from the largest real part of the eigenvalues of the Jacobian matrix $J^*_{PDE}$. Other parameters values are given in Table~\ref{tab:npzd-params}.}
    \label{fig:npzd_population_d1}
\end{center}
\end{figure}

We investigate the stability of the NPZD system in the presence of spatial diffusion. Throughout this study, the parameters listed in Table~\ref{tab:npzd-params} are used, with
$d_2=d_4=10^{-4}\ll d_3=10^{-3}$, since, as shown in the previous section, this parameter regime admits the possibility of a Turing instability. Determining diffusion
coefficients and biological parameters appropriate for specific
ecosystems is an important problem that warrants further investigation, but lies beyond the scope of the present work.

Figure~\ref{fig:npzd_population_d1} shows the maximum linear growth rate of the PDE system as a function of the nutrient diffusion coefficient $d_1$ and the total nitrogen concentration $N_T$ for several vertical wavenumbers
$k=1,2,\ldots,8$. Although half-integer wavenumbers can also occur in this geometry, they are not shown here. Recall that the stability analysis of the corresponding ODE system predicts instability only when $N_T < N_T^* \approx 0.9$. The zero-growth-rate contour initially shifts toward larger values of $N_T$ as $d_1$ increases, reaches a maximum, and then moves back toward smaller values. A Turing instability is possible when this contour crosses the critical threshold $N_T=0.9$, indicating that diffusion has destabilized an equilibrium that is stable in the absence of spatial variation.  The second and third rows demonstrate that several vertical
wavenumbers can generate Turing instabilities. In contrast, the
wavenumbers shown in the bottom row do not produce a Turing
instability, since the zero-growth-rate contour never intersects the critical threshold associated with the ODE system.

To investigate the nonlinear development of the Turing instability, we
consider the parameter regime with $N_T=0.4$, for which the system
admits the equilibrium
\[
[N^*,P^*,Z^*,D^*]^T
\approx
[0.4594,\;
0.0249,\;
0.2694,\;
0.0462]^T.
\]

\begin{figure}[t]
\begin{center}
   \includegraphics[width=\textwidth]{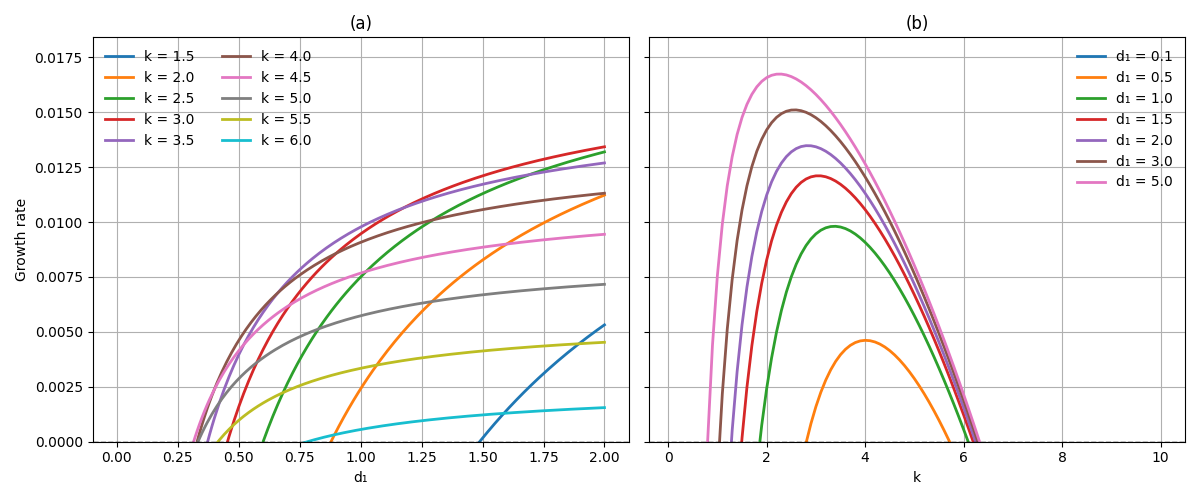}
    \caption{Linear growth rate of the PDE model ~\eqref{eq:NPZD} for various modes and values of the phytoplankton diffusion rate, $d_1$, as computed from the largest real part of the eigenvalues of the Jacobian matrix $J^*_{PDE}$. The rest of the parameters are given in Table~\ref{tab:npzd-params}.}
    \label{fig:npzd_growth_k}
\end{center}
\end{figure}

For this equilibrium, Figure~\ref{fig:npzd_growth_k}(a) shows the
growth rates as a function of the nutrient diffusion coefficient $d_1$
for a range of vertical wavenumbers, while
Figure~\ref{fig:npzd_growth_k}(b) illustrates the corresponding growth
rates as functions of $k$ for several values of $d_1$. The curves in

Figure~\ref{fig:npzd_growth_k}(a) demonstrates that the dependence of
the growth rate on wavenumber is non-monotonic. Small and large wavenumbers are only weakly unstable, while intermediate wavenumbers experience
substantially larger growth rates. The maximum growth rates occur near $k=3$
for $d_1 < 2$, after which the growth rates decrease rapidly as $k$
increases. For large $d_1$ the growth rate of each mode levels off at a maximum value (not shown in plot). Modes with $k>6$ were not included because their growth
rates are negligible, indicating that diffusion suppresses small-scale
perturbations.

Figure~\ref{fig:npzd_growth_k}(b) further reveals that increasing the
nutrient diffusion coefficient initially strengthens the instability.
The growth rates increase with $d_1$ until a parameter regime where the growth rates equilibrate and no longer continue to increase. For the values of $d_1$ shown this is most apparent for $k\ge 4$. This behaviour is
consistent with the classical Turing mechanism, where differential
diffusion destabilizes an otherwise stable spatially homogeneous
equilibrium.

Based on these results, we focus the nonlinear simulations on the
equilibrium above with $d_1=2$. This value lies within a strongly
unstable regime and yields sufficiently large growth rates that the
development of spatial patterns can be observed within practical
simulation times.

In summary, our linear stability analysis predicts that Turing instability can occur in the model \eqref{eq:NPZD} with the parameters given in Table~\ref{tab:npzd-params} when the phytoplankton diffusion constant is sufficiently large, i.e., $d_1>0.3$ where mode $4.5$ becomes unstable in Fig~\ref{fig:npzd_growth_k}(a). We will verify these predictions through numerical simulations of the full model in the next section.

\subsection{Simulations of the Full Model}

In this section, results of simulations of the fully nonlinear model \eqref{eq:NPZD} are presented. We will confirm that Turing instability does occur in the simulations and verify that the linear growth rate in the simulations is as predicted by the linear stability analysis. The initial conditions are set up as perturbations from the equilibrium:

\begin{align}
\hspace{-.2in}
    [N, P, Z, D](x, 0) &= [N^*, P^*, Z^*, D^*]+[W_1, W_2, W_3, W_4 ](x). \label{eq:ICn}
\end{align}
We consider two choices for the perturbation functions, $W_j(x)$: a single mode and a random mode from a normal distribution. The former is used to verify the linear growth, while the latter allows us to study the effect of all the unstable modes on the evolution of the model.

To facilitate comparison between the numerical simulations and the linear stability analysis, we present the results as perturbations from the equilibrium solution, i.e., we plot the difference

\[ [N,P,Z,D](x,t)-[N^*,P^*,Z^*,D^*]. \]

\subsubsection{Single mode with $k=3$}

\begin{figure}[!tp]
\begin{center}
   \includegraphics[width=\textwidth]{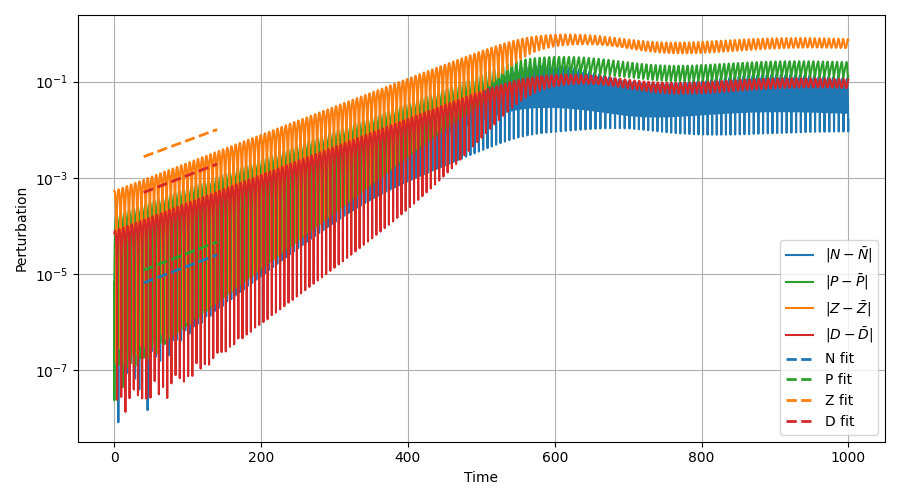}
    \caption{Plots of the norm of the perturbation fields in the numerical simulations where the initial condition has a single wavenumber, $k=3$.}
    \label{fig:simulation_keq3_growth}
\end{center}
\end{figure}

Figure~\ref{fig:npzd_growth_k} shows that, for $d_1 = 2$, the largest linear growth rate of the PDE system occurs for the mode $k=3$. Accordingly, we choose the initial perturbation in the form
\begin{align}
W_j(x) = \cos(3\pi x)\, v_{3j},
\label{eq:ICnk3}
\end{align}
where ${\bf v}_3 = (v_{31}, v_{32}, v_{33}, v_{34})$ is the eigenvector corresponding to the unstable eigenvalue (i.e., the eigenvalue with positive real part) for $k=3$ shown in Figure~\ref{fig:npzd_growth_k}.

\begin{table}
\centering
\caption{Measured exponential growth rates obtained from the nonlinear simulations. 
of Figures~\ref{fig:simulation_keq3_growth} and \ref{fig:simulation_rand_growth}.
The growth rate predicted by the linear stability analysis is $0.0134$.}
\label{tab:growth_rates}
\begin{tabular}{lcc}
\toprule
Component &
Single mode ($k=3$) &
Random initial condition \\
\midrule
Nutrients (N)     & 0.0134 & 0.0119 \\
Phytoplankton (P) & 0.0135 & 0.0126 \\
Zooplankton (Z)   & 0.0130 & 0.0129 \\
Detritus (D)      & 0.0136 & 0.0270 \\
\bottomrule
\end{tabular}
\end{table}

We present several diagnostics from this simulation. Figure~\ref{fig:simulation_keq3_growth} shows the norm of the perturbation from the equilibrium as a function of time. Four best-fit lines are included to estimate the exponential growth rates of the different components. These lines are nearly parallel, with slopes ranging between $0.0134$ and $0.0136$, in excellent agreement with the growth rate predicted by the linear stability analysis, namely $0.0134$. See Table~\ref{tab:growth_rates}.  During the initial phase of exponential growth, oscillations are superimposed on the growing signal; the fitted slopes therefore characterize the growth of the envelope rather than the instantaneous fluctuations. Nonlinear saturation occurs at approximately $t=600$, after which the curves level off while retaining a small oscillatory component. Overall, the close agreement between the measured and predicted growth rates demonstrates that the linear theory provides an accurate description of the early-time evolution of this simulation.

\begin{figure}
\begin{center}
   \includegraphics[width=\textwidth]{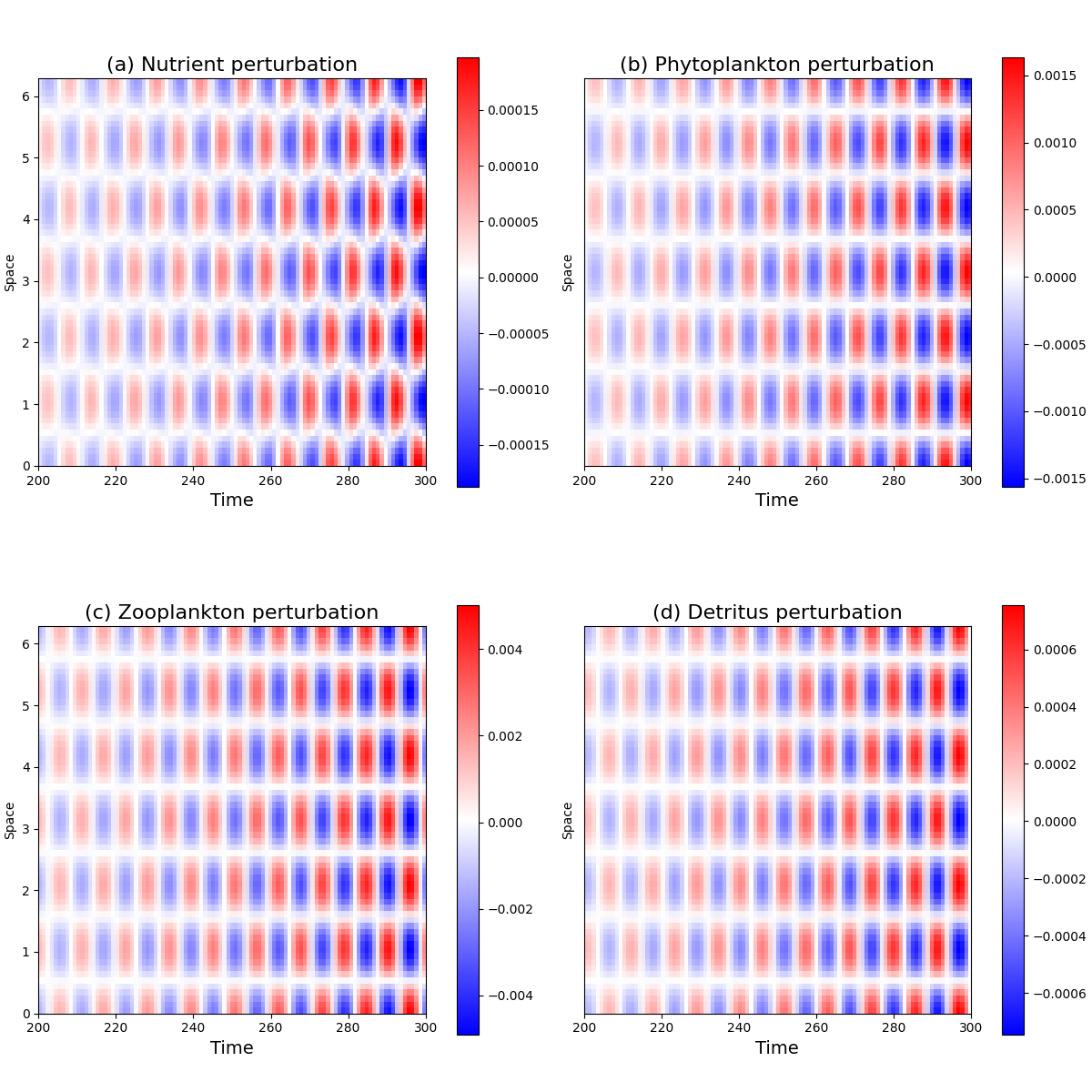}
    \caption{Plots of the perturbations of the four fields, $N, P, Z$, and $D$, in the early stages of the  numerical simulations where the initial condition has a single wavenumber, $k=3$.}
    \label{fig:simulation_keq3_early_fields}
\end{center}
\end{figure}

Figure~\ref{fig:simulation_keq3_early_fields} presents Hovmöller diagrams (where time and space are the horizontal and vertical axes, respectively) of the four state variables at an early stage of the simulation, well within the regime of exponential growth. The characteristic three-wave structure in the spatial domain is clearly visible in each field and is consistent with the dominant unstable mode predicted by the linear stability analysis. As expected, the qualitative form of the solution remains essentially unchanged throughout this growth phase. The primary evolution is an increase in amplitude, while the spatial structure of each field is preserved. This behavior is characteristic of a system whose dynamics are dominated by a single exponentially growing eigenmode.

\begin{figure}
\begin{center}
   \includegraphics[width=\textwidth]{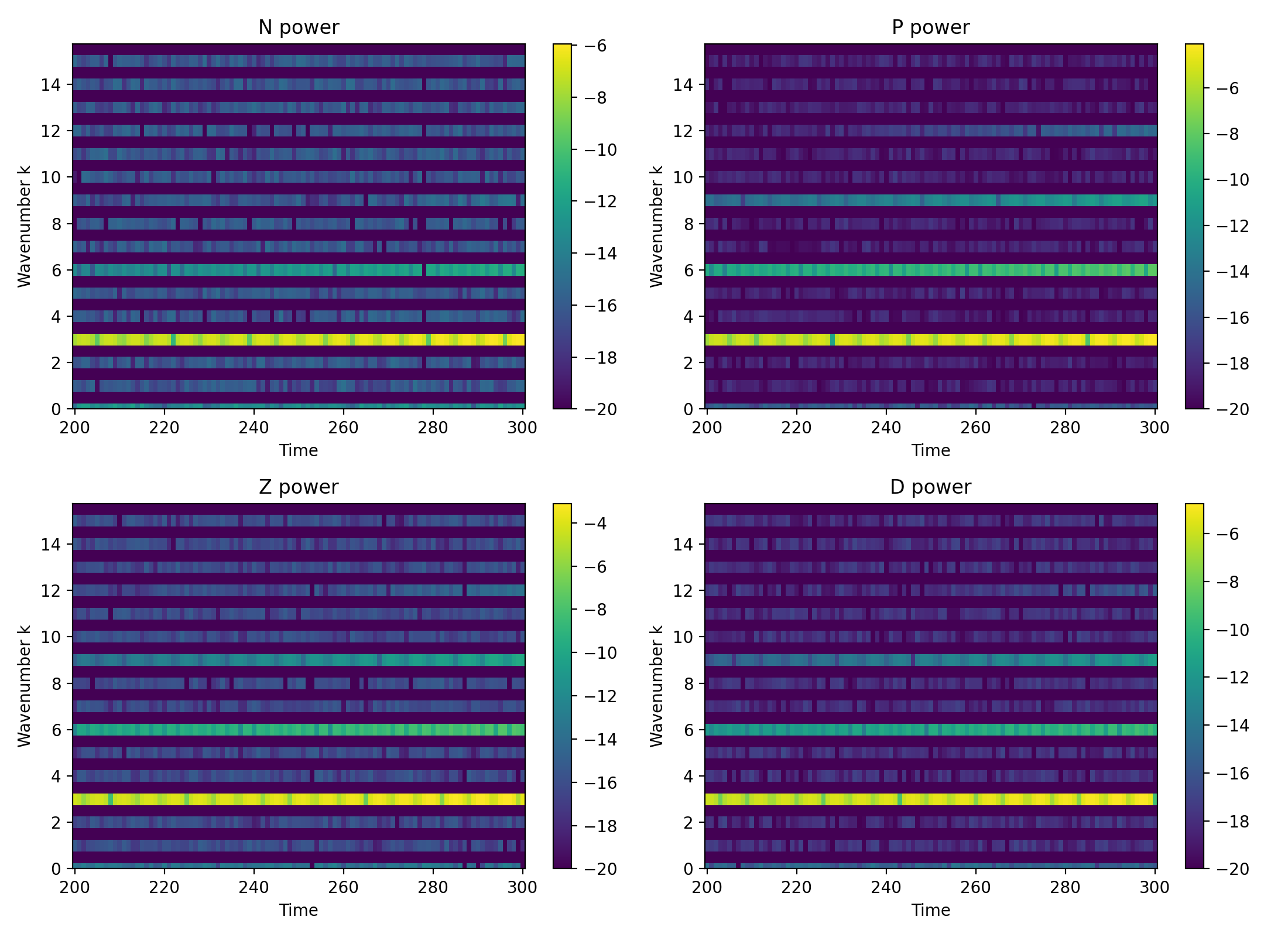}
    \caption{Plots of the spectra of the perturbations of the four fields, $N, P, Z$, and $D$, in the early stages of the  numerical simulations where the initial condition has a single wavenumber, $k=3$.}
    \label{fig:simulation_keq3_early_spectra}
\end{center}
\end{figure}

Figure~\ref{fig:simulation_keq3_early_spectra} shows the power spectra of the four fields over the same time interval considered in Figure~\ref{fig:simulation_keq3_early_fields}. In each case, a pronounced peak is observed at $k=3$, confirming that the solution is dominated by the most unstable mode predicted by the linear stability analysis. The spectra also reveal the presence of weaker higher-wavenumber components. In particular, secondary peaks are visible at $k=6$ and $k=9$, corresponding to the second and third harmonics of the dominant mode. The amplitudes of these harmonics decrease with increasing wavenumber, indicating that the solution remains strongly concentrated at the fundamental mode. Although all wavenumbers are present at some level due to the nonlinear dynamics and numerical representation of the fields, only a small number of harmonics rise appreciably above the spectral background. This demonstrates that the evolving solution retains a highly coherent structure despite the emergence of weak nonlinear interactions.

\begin{figure}
\begin{center}
   \includegraphics[width=\textwidth]{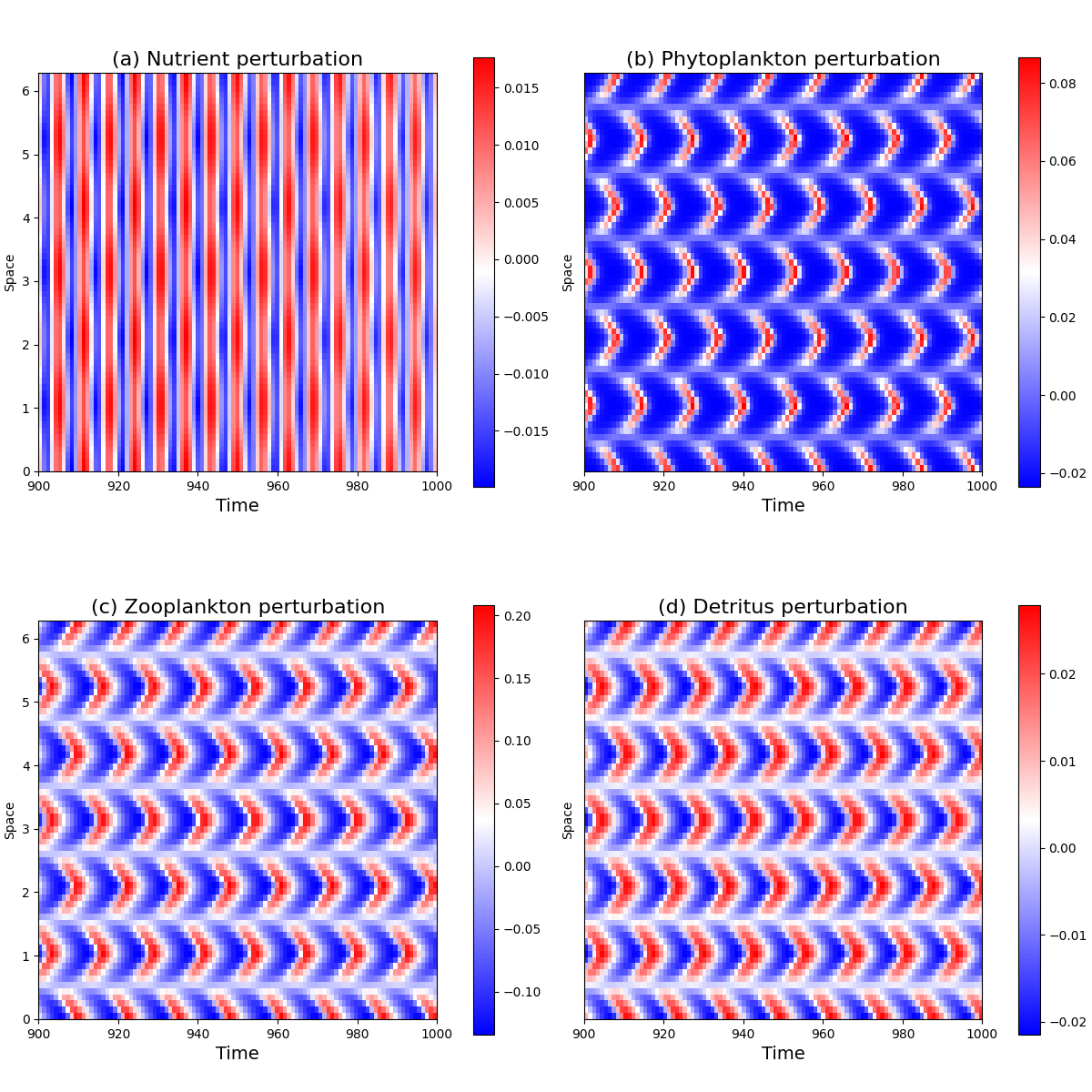}
    \caption{Plots of the perturbations of the four fields, $N, P, Z$, and $D$, in the later stages of the  numerical simulations where the initial condition has a single wavenumber, $k=3$.}
    \label{fig:simulation_keq3_late_fields}
\end{center}
\end{figure}

Figure~\ref{fig:simulation_keq3_late_fields} presents the same fields as Figure~\ref{fig:simulation_keq3_early_fields}, but at later times after nonlinear equilibration has occurred. In this regime, nonlinear interactions have altered the solution and led to substantial modifications of the field structure. Despite these changes, the dominant three-wave pattern remains clearly identifiable throughout the simulation.

The nutrient field develops a pronounced spatially averaged component that oscillates in time, making the underlying structure $k=3$ less apparent than in the other variables. In contrast, the phytoplankton, zooplankton, and detritus fields continue to exhibit a strong three-wave pattern in the nonlinear regime, although their temporal evolution differs markedly from that observed during the linear growth phase. In particular, these fields develop a distinctive chevron-shaped structure in the time--space diagrams. The solution remains qualitatively similar at all later times, indicating that the system has reached a persistent nonlinear state characterized by oscillations about the equilibrated configuration.

\begin{figure}
\begin{center}
   \includegraphics[width=\textwidth]{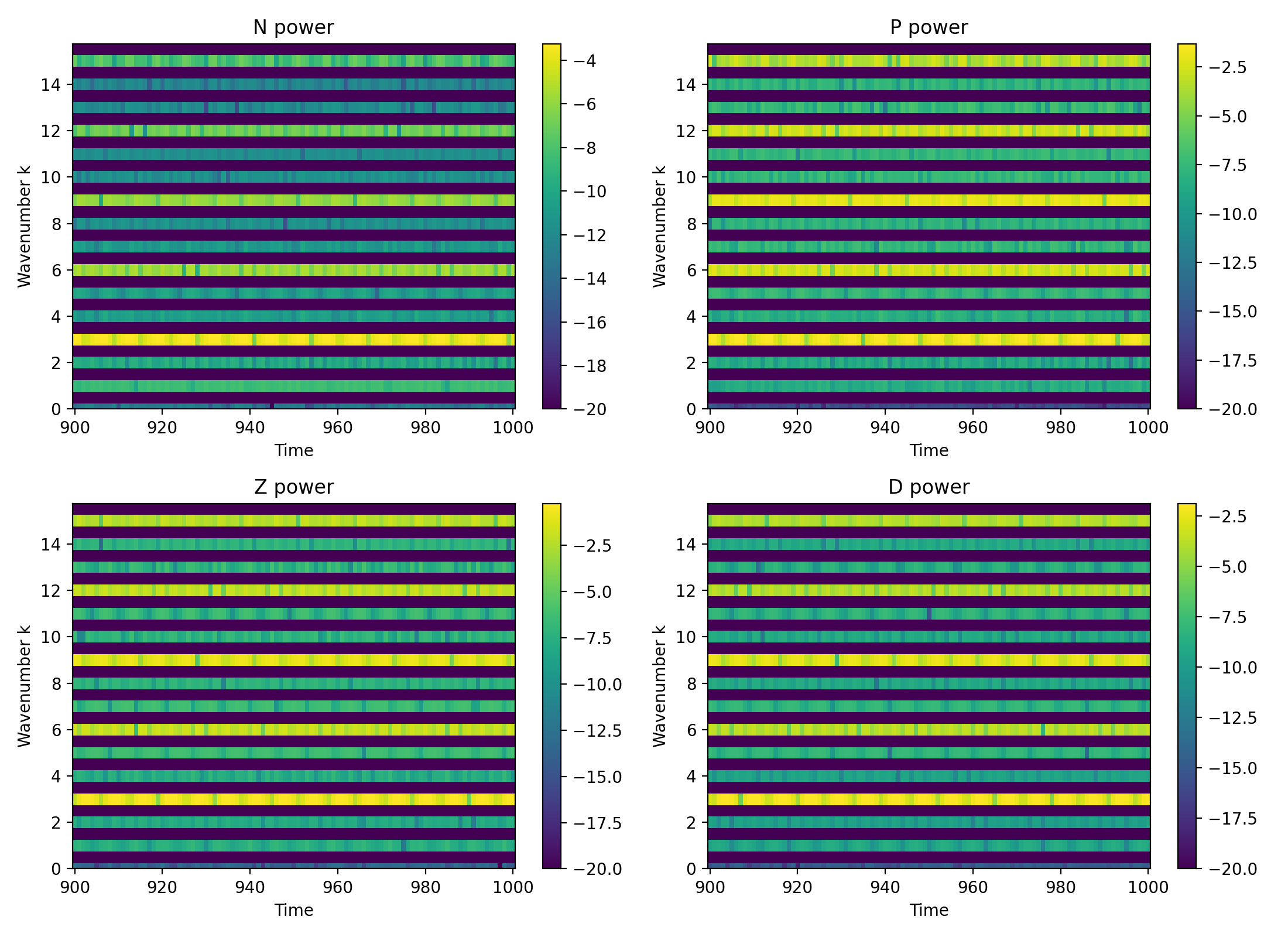}
    \caption{Plots of the spectra of the perturbations of the four fields, $N, P, Z$, in the later stages of the  numerical simulations where the initial condition has a single wavenumber, $k=3$.}
    \label{fig:simulation_keq3_late_spectra}
\end{center}
\end{figure}

Finally, Figure~\ref{fig:simulation_keq3_late_spectra} shows the power spectra corresponding to the nonlinear regime illustrated in Figure~\ref{fig:simulation_keq3_late_fields}. As in the early stages of the simulation, the dominant spectral peak remains at $k=3$ for all four fields, confirming that the primary pattern is still governed by the most unstable mode. However, the higher-wavenumber components are significantly more energetic than during the linear growth phase. In particular, the harmonics of the fundamental mode attain amplitudes that are only modestly smaller than that of the $k=3$ component. This behavior is evident in all four fields and reflects the increasing importance of nonlinear interactions after equilibration. In addition, non-harmonic wavenumbers become more pronounced, although they remain weaker than the dominant mode and its harmonics. The resulting spectra indicate that solution has been redistributed across a broader range of spatial scales while preserving the underlying three-wave structure.

Together, these results provide strong validation of the linear stability analysis and clearly demonstrate the onset of a Turing instability in the PDE system. The simulations show that the growth rate and spatial structure predicted by the linear theory accurately describe the initial evolution, while nonlinear effects eventually lead to a saturated state containing a richer spectrum of spatial scales. Because the initial condition consists of a single unstable mode, the opportunities for energy transfer among scales are relatively limited. In the next subsection, we revisit the same parameter regime using a random initial perturbation, allowing a broader range of modes to interact and providing a more complete picture of the nonlinear pattern-forming dynamics.

\subsubsection{Random initial perturbation}

\begin{figure}
\begin{center}
   \includegraphics[width=\textwidth]{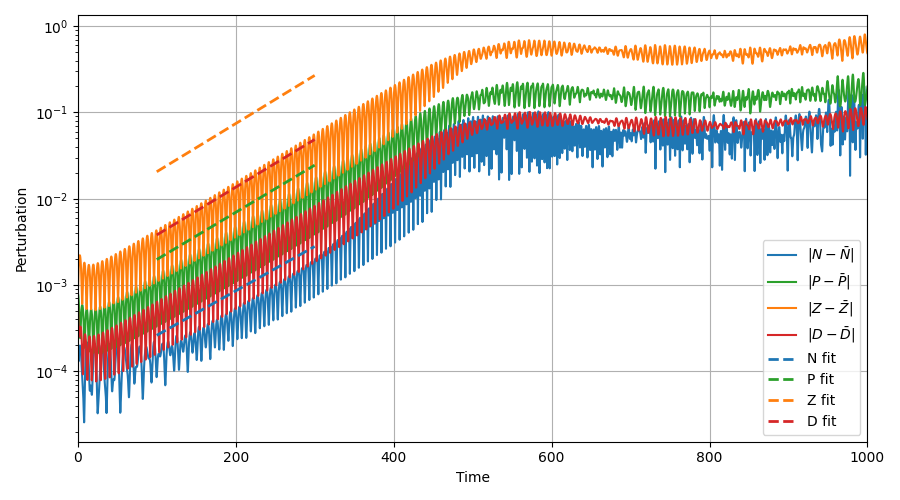}
    \caption{Plots of the norm of the perturbation fields in the numerical simulations where the initial condition is random.}
    \label{fig:simulation_rand_growth}
\end{center}
\end{figure}

We now examine the numerical simulation in which the initial conditions for all four fields are generated from independent Gaussian random perturbations.  To do this we pick the initial conditions in the form \eqref{eq:ICn} with $W_j(x)$ chosen randomly from a normal distribution as represented in equation~\eqref{eq:IC_random},
\begin{align} 
W_j(x) &= \epsilon_j(x), \quad \epsilon_j(x) \sim \mathcal{N}(0, \sigma^2)
\label{eq:IC_random}
\end{align} 
where $\epsilon_j$ is a random variable which follows a normal distribution with standard deviation $\sigma$. 

Figure~\ref{fig:simulation_rand_growth} shows the norms of the perturbation fields as functions of time, together with dashed lines representing least-squares linear fits during the growth phase. As in the previous experiment, the solution exhibits a clear regime of exponential growth, and the fitted growth rates are similar across all four components. The estimated slopes lie between $0.012$ and $0.013$, which are slightly smaller than those obtained in the single-mode simulation. See Table~\ref{tab:growth_rates} for a comparison. This difference is expected because the random initial condition excites a broad range of wavenumbers, including many modes that are less unstable than the dominant mode predicted by the linear stability analysis.  Recall that Figure \ref{fig:npzd_population_d1} showed that all wavenumbers from $k=1.5$ to $k=6$ were linearly unstable and hence we expect they will all be present in this simulation.

To reduce the computational cost of the experiment, the initial perturbation amplitude was chosen to be larger than in the previous simulation. Consequently, nonlinear equilibration occurs earlier, at approximately $t=500$. After saturation, the fields exhibit oscillatory behavior that is qualitatively similar to that observed in the single-mode case. Overall, the results again demonstrate the presence of an initial exponential growth phase followed by nonlinear equilibration, while highlighting the effects of initializing the system with a spectrum of spatial scales rather than a single unstable mode.

\begin{figure}
\begin{center}
   \includegraphics[width=\textwidth]{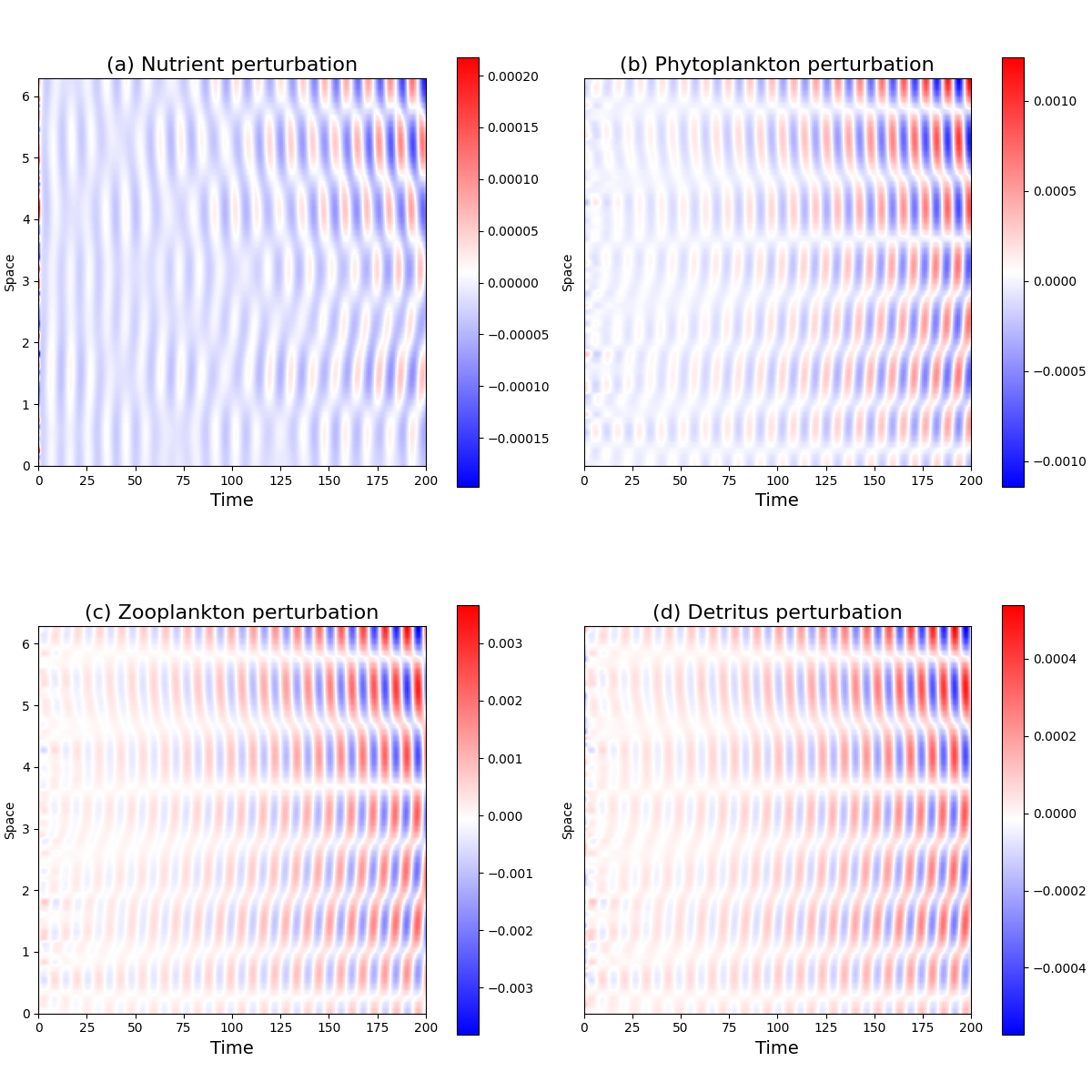}
    \caption{Plots of the perturbations of the four fields, $N, P, Z$, and $D$, in the early stages of the  numerical simulations where the initial condition is random.}
    \label{fig:simulation_rand_early_fields}
\end{center}
\end{figure}

Figure~\ref{fig:simulation_rand_early_fields} shows Hovmöller diagrams of the four components during the first $200$ time units of the simulation. At these early times, no dominant spatial mode is immediately apparent, reflecting the broad spectrum of wavenumbers excited by the random initial conditions. However, as the simulation evolves, a coherent wave pattern gradually emerges. By the end of the displayed interval, the solution is clearly dominated by a structure with a wavelength corresponding approximately to $k=3$, consistent with the prediction of the linear stability analysis.

It is important to emphasize, however, that the solution is not composed solely of this dominant mode. The linear theory predicts that multiple wavenumbers are unstable, and all of these are initially present due to the random perturbations. Many of these additional modes remain difficult to identify directly from the Hovmöller diagrams, either because their amplitudes are comparatively small or because they are superimposed on the dominant pattern. To quantify the relative importance of the various spatial scales and better assess the modal content of the solution, we therefore turn to an analysis of the power spectra.

\begin{figure}
\begin{center}
   \includegraphics[width=\textwidth]{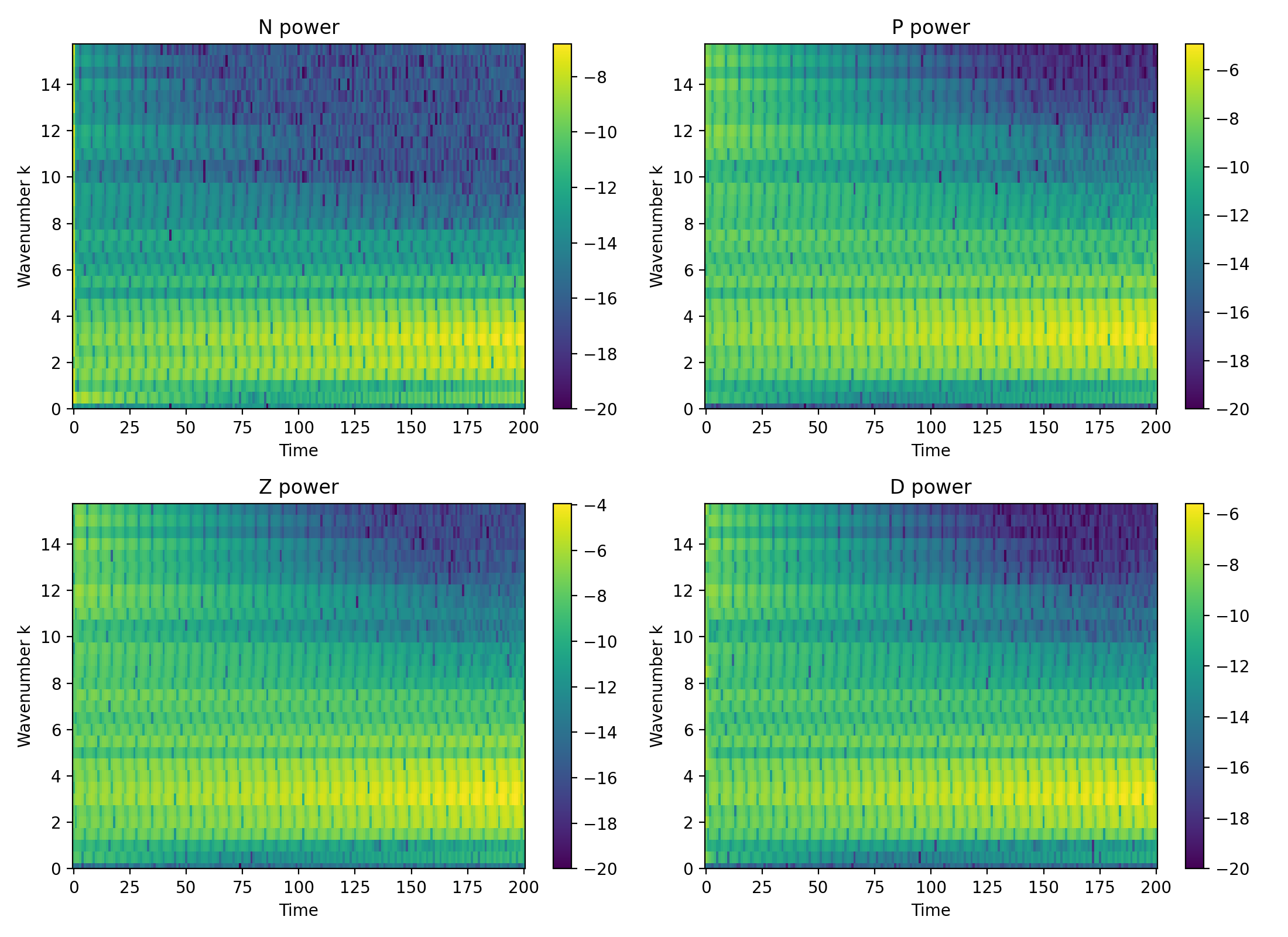}
    \caption{Plots of the spectra of perturbations of the four fields $N, P, Z$, and $D$, in the early stages of the  numerical simulations where the initial condition is random.}
    \label{fig:simulation_rand_early_spectra}
\end{center}
\end{figure}

Figure~\ref{fig:simulation_rand_early_spectra} shows the corresponding power spectra during the early stages of the simulation. Initially, the spectral energy in all four fields is distributed relatively uniformly across a broad range of wavenumbers, reflecting the random nature of the initial perturbations. As the simulation evolves, however, a distinct concentration of energy emerges near $k=3$. By approximately $t=100$, this dominant spectral feature is clearly visible in all four fields, consistent with the predictions of the linear stability analysis.  In particular, the modal structure of the solution at $t=200$ is very similar to what is predicted from the $d_1=2$ curve in Figure \ref{fig:npzd_population_d1} in that the maximum brightness occurs near $k=3$ and then then decreases with larger and smaller $k$, with very little brightness for $k>6$ or $k<1$. Thus the strength of each mode in the solution correlates with the size of the linear growth rate.

Unlike the single-mode experiment, where the spectrum was concentrated in a narrow band centered on $k=3$, the present spectra exhibit a much broader peak. This broader structure arises because the random initial condition excites many spatial modes, several of which are linearly unstable. Indeed, the linear theory predicts instability over a range of wavenumbers rather than at a single isolated mode. Consequently, multiple unstable modes grow simultaneously and contribute to the evolving spectral distribution. Nevertheless, the simulation remains in an early stage of development, and the spectra shown here do not yet reveal the fully equilibrated state of the system. To understand the long-term outcome and the effects of nonlinear interactions, we next examine the solution at later times.

\begin{figure}
\begin{center}
   \includegraphics[width=\textwidth]{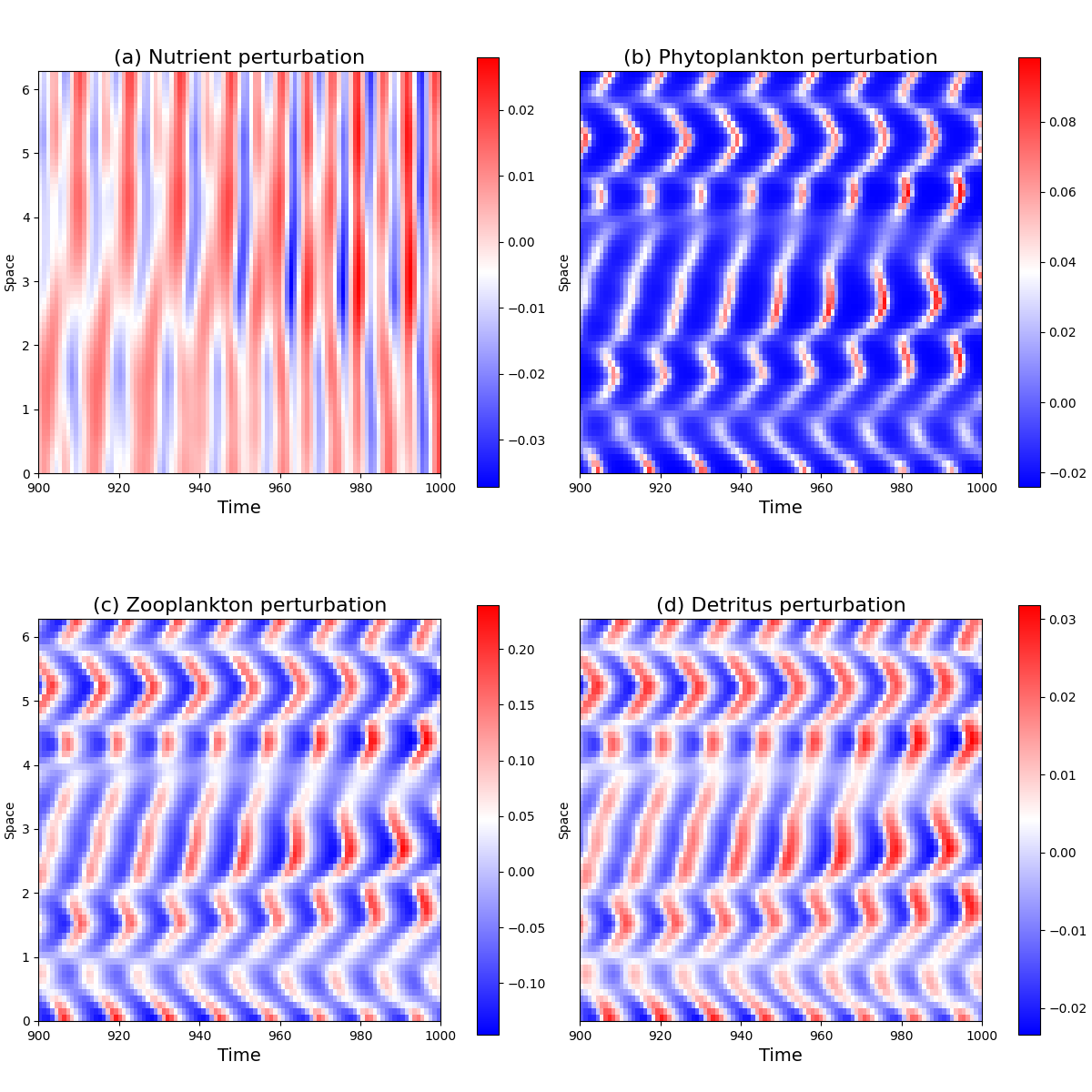}
    \caption{Plots of the perturbations of the four fields, $N, P, Z$, and $D$, in the later stages of the  numerical simulations where the initial condition is random.}
    \label{fig:simulation_rand_late_fields}
\end{center}
\end{figure}

Figure~\ref{fig:simulation_rand_late_fields} shows the solution at later times, well after nonlinear saturation has been reached. While some remnants of the chevron-shaped structures observed in the single-mode experiment remain visible, the overall dynamics are considerably more complex. The nutrient field retains a degree of vertical uniformity similar to that seen previously, but now also exhibits a pronounced depth-dependent structure. The phytoplankton, zooplankton, and detritus fields display patterns that resemble propagating or traveling waves, indicating richer spatiotemporal behavior in the nonlinear regime.

The zooplankton and detritus structures are generally broader and more diffuse than those of the phytoplankton, suggesting differences in the characteristic spatial scales associated with these variables. This is interesting as the diffusion constants for phytoplankton and detritus are equal.  More importantly, the fields exhibit a substantial degree of irregularity that was absent in the single-mode simulation. This increased complexity arises from the random initial conditions, which excite a wide range of unstable modes that subsequently interact through nonlinear processes. As a result, the equilibrated state does not settle into a simple periodic pattern, but instead displays persistent variability while still retaining signatures of the dominant Turing-instability dynamics.

\begin{figure}[t]
\begin{center}
   \includegraphics[width=\textwidth]{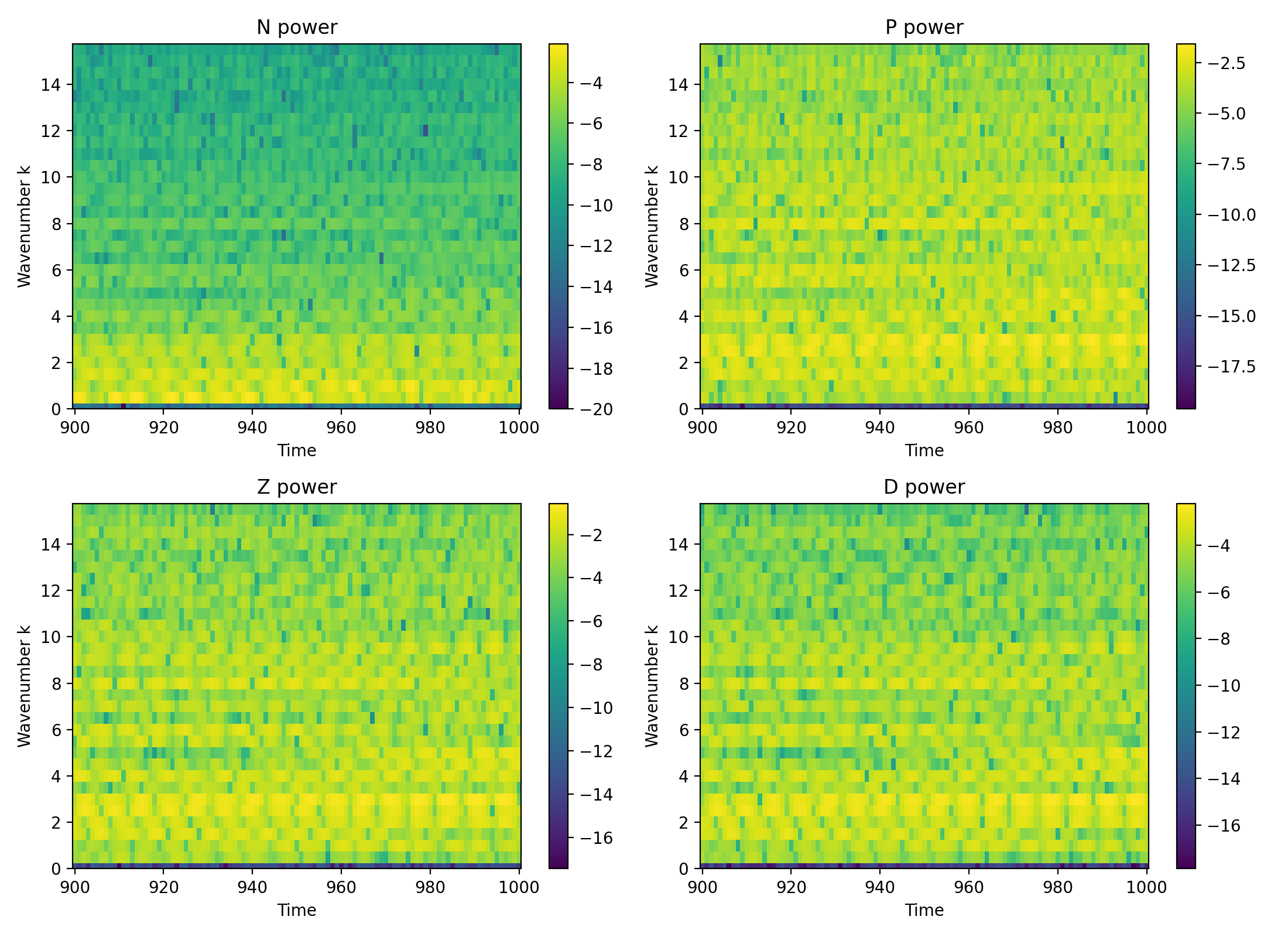}
    \caption{Plots of the spectra of the perturbations of the four fields, $N, P, Z$, and $D$, in the later stages of the  numerical simulations where the initial condition is random.}
    \label{fig:simulation_rand_late_spectra}
\end{center}
\end{figure}

Finally, Figure~\ref{fig:simulation_rand_late_spectra} shows the power spectra after nonlinear equilibration has been established. These spectra differ markedly from those observed in the single-mode experiment. In the phytoplankton, zooplankton, and detritus fields, the spectral energy is distributed over a broad range of wavenumbers, indicating substantial transfer of energy across spatial scales. The nutrient field exhibits a somewhat narrower distribution, although it too contains energy over a range of modes. In all four cases, the largest concentrations of spectral power occur at wavenumbers $k \leq 3$, reflecting the influence of the most unstable modes identified by the linear stability analysis. However, significant energy is also present at higher wavenumbers, demonstrating the increasingly important role of nonlinear interactions in shaping the equilibrated state.

This result provides important insight into the potential impact of Turing instabilities on biological systems. As demonstrated throughout this section, the linear theory accurately predicts the initial growth rates and identifies the spatial scales that first emerge from perturbations of the equilibrium state, even when the initial conditions are random. However, the linear analysis alone provides little information about the eventual nonlinear state. Both the Hovmöller diagrams and the power spectra reveal that the equilibrated solution contains a rich collection of spatial structures spanning a wide range of scales. Predicting the detailed form of these structures is challenging because of the complex and potentially chaotic interactions among unstable modes. Nevertheless, the simulations clearly show that Turing instabilities can generate biologically relevant variability across many spatial scales, extending far beyond the patterns anticipated from the dominant linear modes alone.

\section{Conclusion}
\label{sec:conclusions}

In this work, we have investigated a conservative NPZD model as an idealized representation of a planktonic ecosystem consisting of four interacting components: nutrients, phytoplankton, zooplankton, and detritus. We identified spatially homogeneous equilibrium solutions in which both phytoplankton and zooplankton persist and examined their stability in the context of both the corresponding ODE system and the reaction--diffusion PDE system. Our analysis revealed a range of parameter values for which the homogeneous equilibrium is stable in the absence of diffusion but becomes unstable when diffusion is introduced. This behavior is characteristic of a Turing instability, whereby diffusion destabilizes a uniform state and drives the emergence of spatial patterns with preferred wavelengths. The predictions of the linear stability analysis were found to be in excellent agreement with the nonlinear simulations, both in terms of the observed growth rates and the spatial structures that developed.  

The simulation initialized with a single wavenumber reproduced the predictions of the linear stability analysis with remarkable accuracy, both in terms of the exponential growth rates and the resulting spatial structure. Following nonlinear saturation, the shape of the solution evolved, but the dominant wavenumber remained unchanged. This behavior contrasts sharply with the simulation initialized with random perturbations, where the dynamics were considerably more complex. In that case, multiple unstable wavenumbers grew simultaneously during the early stages of the evolution, leading ultimately to a much broader power spectrum. These results highlight the effectiveness of Turing instabilities as a mechanism for generating spatial patterns across a wide range of length scales and demonstrate how the structure of the initial perturbation can strongly influence the subsequent nonlinear dynamics.

The Turing instabilities observed in our study occurred when the nutrient diffusion coefficient was several orders of magnitude larger than those of the other components. While we do not claim that such parameter values are representative of all biological systems, they nevertheless provide insight into the mechanisms by which diffusion can promote spatial pattern formation. Rapid nutrient diffusion allows nutrients to be redistributed throughout the domain much more quickly than phytoplankton, zooplankton, or detritus. Regions with low phytoplankton biomass can therefore receive nutrients on relatively short timescales, promoting local phytoplankton growth. This enhanced growth subsequently supports increases in zooplankton biomass and, through mortality and recycling, detritus production. Under appropriate conditions, these interactions can amplify spatial variations and lead to the formation of persistent patterns. Although the strongest instabilities identified here required a large disparity in diffusivities, it is possible that Turing instabilities may also occur for more modest diffusion ratios. Determining the range of biologically realistic parameter values that support such behavior remains an important topic for future investigation.

Our work extends prior analyses of Turing instability for two- and three-component systems \cite{murray2007mathematical} to a four-component, conservative, diffusive NPZD model. We show that different diffusion coefficients induce Turing instability in contrast to previous work on a similar model that showed that equal diffusion coefficients improve stability \cite{tao2021dynamics}. Thus, our work further supports the generally accepted belief that differences between diffusion coefficients are necessary for Turing instability. 

One promising direction for future research is to investigate more complex plankton ecosystem models in order to determine how robust and widespread Turing instabilities are in biologically realistic settings. The NPZD formulation considered here is highly idealized, and the inclusion of additional ecological processes, such as multiple nutrient classes, higher trophic levels, or environmental forcing, may significantly influence the onset and evolution of pattern formation. Understanding whether diffusion-driven instabilities persist in these more sophisticated models would provide valuable insight into the generality of the mechanisms identified in this study.  As well, including advection which is known to be very important because of sinking, can and should be included in future studies.

A second important avenue is to examine parameter regimes representative of specific marine ecosystems. By combining field observations with model analysis, it may be possible to identify locations and environmental conditions under which Turing instabilities could occur in nature. Such investigations would help bridge the gap between idealized theoretical models and observed plankton distributions, and could shed light on the role of diffusion-driven pattern formation in shaping marine ecosystem structure and variability.

\bibliographystyle{plain}
\bibliography{Turing_paper}

@article{gentleman2002chronology,
  title={A chronology of plankton dynamics in silico: how computer models have been used to study marine ecosystems},
  author={Gentleman, Wendy},
  journal={Hydrobiologia},
  volume={480},
  number={1},
  pages={69--85},
  year={2002},
  publisher={Springer}
}

@article{taniguchi2014planktonic,
  title={Planktonic biomass size spectra: an emergent property of size-dependent physiological rates, food web dynamics, and nutrient regimes},
  author={Taniguchi, Darcy AA and Franks, Peter JS and Poulin, Francis J},
  journal={Marine Ecology Progress Series},
  volume={514},
  pages={13--33},
  year={2014}
}

@article{beretta1990stability,
  title={Stability in chemostat equations with delayed nutrient recycling},
  author={Beretta, Edoardo and Bischi, GI and Solimano, Fortunata},
  journal={Journal of Mathematical Biology},
  volume={28},
  number={1},
  pages={99--111},
  year={1990},
  publisher={Springer}
}

@article{kloosterman2014closed,
  title={A closed {NPZ} model with delayed nutrient recycling},
  author={Kloosterman, Matthew and Campbell, Sue Ann and Poulin, Francis J},
  journal={Journal of Mathematical Biology},
  volume={68},
  number={4},
  pages={815--850},
  year={2014},
  publisher={Springer}
}

@article{tao2021dynamics,
  title={Dynamics of a diffusive nutrient-phytoplankton-zooplankton model with spatio-temporal delay},
  author={Tao, Yiwen and Campbell, Sue Ann and Poulin, Francis J},
  journal={SIAM Journal on Applied Mathematics},
  volume={81},
  number={6},
  pages={2405--2432},
  year={2021},
  publisher={SIAM}
}

@book{murray2007mathematical,
  title={Mathematical biology: I. An introduction},
  author={Murray, James D},
  volume={17},
  year={2007},
  publisher={Springer Science \& Business Media}
}

@book{murraymathematical2,
  title={Mathematical Biology: II: Spatial Models and Biomedical Applications},
  author={Murray, James D},
 year={2003},
  publisher={Springer Science \& Business Media}
}

@article{svansson1996diffusion,
  title={A diffusion model for the primary production of phytoplankton},
  author={Svansson, Artur},
  journal={Deep Sea Research Part II: Topical Studies in Oceanography},
  volume={43},
  number={1},
  pages={37--46},
  year={1996},
  publisher={Elsevier}
}

@article{he1998global,
  title={Global stability in chemostat-type plankton models with delayed nutrient recycling},
  author={He, Xue-Zhong and Ruan, Shigui},
  journal={Journal of Mathematical Biology},
  volume={37},
  number={3},
  pages={253--271},
  year={1998},
  publisher={Springer}
}

@article{ruan1995effect,
  title={The effect of delays on stability and persistence in plankton models},
  author={Ruan, Shigui},
  journal={Nonlinear Analysis: Theory, Methods \& Applications},
  volume={24},
  number={4},
  pages={575--585},
  year={1995},
  publisher={Pergamon}
}

@article{ruan1998turing,
  title={Turing instability and travelling waves in diffusive plankton models with delayed nutrient recycling},
  author={Ruan, SHIGUI},
  journal={IMA Journal of Applied Mathematics},
  volume={61},
  number={1},
  pages={15--32},
  year={1998},
  publisher={Oxford University Press}
}

@article{ruan2001oscillations,
  title={Oscillations in plankton models with nutrient recycling},
  author={Ruan, Shigui},
  journal={Journal of Theoretical Biology},
  volume={208},
  number={1},
  pages={15--26},
  year={2001},
  publisher={Elsevier}
}

@book{ogata2010modern,
  title={Modern control engineering},
  author={Ogata, Katsuhiko},
  year={2010},
  publisher={Prentice hall}
}

@article{du2010nonlocal,
  title={On a nonlocal reaction--diffusion--advection equation modelling phytoplankton dynamics},
  author={Du, Yihong and Mei, Linfeng},
  journal={Nonlinearity},
  volume={24},
  number={1},
  pages={319},
  year={2010},
  publisher={IOP Publishing}
}

@article{franks1986prediction,
  title={Prediction of phytoplankton growth in response to the frictional decay of a warm-core ring},
  author={Franks, Peter JS and Wroblewski, Joe S and Flierl, Glenn R},
  journal={Journal of Geophysical Research: Oceans},
  volume={91},
  number={C6},
  pages={7603--7610},
  year={1986},
  publisher={Wiley Online Library}
}

@article{franks2002npz,
  title={{NPZ} models of plankton dynamics: their construction, coupling to physics, and application},
  author={Franks, Peter JS},
  journal={Journal of Oceanography},
  volume={58},
  number={2},
  pages={379--387},
  year={2002},
  publisher={Springer}
}

@article{kmet1996material,
  title={Material recycling in a closed aquatic ecosystem. {I}. {N}itrogen transformation cycle and preferential utilization of ammonium to nitrate by phytoplankton as an optimal control problem},
  author={Kmet', Tibor},
  journal={Bulletin of Mathematical Biology},
  volume={58},
  number={5},
  pages={957--982},
  year={1996},
  publisher={Springer}
}

@inproceedings{lu2021turing,
  title={Turing instability analysis of marine planktonic ecosystem under the influence of spatial heterogeneity},
  author={Lu, Yunxiang and Xiao, Min and Chen, Gong and Tao, Binbin and Chen, Shi and Liu, Jiaxuan},
  booktitle={2021 33rd Chinese Control and Decision Conference (CCDC)},
  pages={5333--5338},
  year={2021},
  organization={IEEE}
}

@article{poulin2010size,
  title={Size-structured planktonic ecosystems: constraints, controls and assembly instructions},
  author={Poulin, Francis J and Franks, Peter JS},
  journal={Journal of Plankton Research},
  volume={32},
  number={8},
  pages={1121--1130},
  year={2010},
  publisher={Oxford University Press}
}

@article{Julia-2017,
    title={Julia: A fresh approach to numerical computing},
    author={Bezanson, Jeff and Edelman, Alan and Karpinski, Stefan and Shah, Viral B},
    journal={SIAM {R}eview},
    volume={59},
    number={1},
    pages={65--98},
    year={2017},
    publisher={SIAM},
    doi={10.1137/141000671}
}

@article{OceananigansJOSS,
  doi = {10.21105/joss.02018},
  year = {2020},
  publisher = {The Open Journal},
  volume = {5},
  number = {53},
  pages = {2018},
  author = {Ali Ramadhan and Gregory LeClaire Wagner and Chris Hill and Jean-Michel Campin and Valentin Churavy and Tim Besard and Andre Souza and Alan Edelman and Raffaele Ferrari and John Marshall},
  title = {Oceananigans.jl: Fast and friendly geophysical fluid dynamics on {GPU}s},
  journal = {Journal of Open Source Software}
}

@techreport{davis1994biological,
  title={Biological/Physical Modeling of Upper Ocean Processes.},
  author={Davis, Cabell S and Steele, John H},
  year={1994},
institution={Woods Hole Oceanographic Institution}
}

@article{Edwards2001,
    author = {Edwards, Andrew M},
    title = {Adding Detritus to a Nutrient–Phytoplankton–Zooplankton Model: A Dynamical Systems Approach},
    journal = {Journal of Plankton Research},
    volume = {23},
    number = {4},
    pages = {389-413},
    year = {2001},
    month = {04},
    issn = {0142-7873},
    doi = {10.1093/plankt/23.4.389}
}

@article{courant1967partial,
  title={On the partial difference equations of mathematical physics},
  author={Courant, Richard and Friedrichs, Kurt and Lewy, Hans},
  journal={IBM Journal of Research and Development},
  volume={11},
  number={2},
  pages={215--234},
  year={1967},
  publisher={IBM}
}

@article{armbruster2012merging,
  title={Merging mythology and morphology: the multifaceted lifestyle of Proteus mirabilis},
  author={Armbruster, Chelsie E and Mobley, Harry LT},
  journal={Nature Reviews Microbiology},
  volume={10},
  number={11},
  pages={743--754},
  year={2012},
  publisher={Nature Publishing Group UK London}
}

@article{krause2021isolating,
  title={Isolating patterns in open reaction--diffusion systems},
  author={Krause, Andrew L and Klika, V{\'a}clav and Maini, Philip K and Headon, Denis and Gaffney, Eamonn A},
  journal={Bulletin of Mathematical Biology},
  volume={83},
  number={7},
  pages={82},
  year={2021},
  publisher={Springer}
}

@article{marciniak2017instability,
  title={Instability of Turing patterns in reaction-diffusion-{ODE} systems},
  author={Marciniak-Czochra, Anna and Karch, Grzegorz and Suzuki, Kanako},
  journal={Journal of Mathematical Biology},
  volume={74},
  number={3},
  pages={583--618},
  year={2017},
  publisher={Springer}
}

@article{shao2022turing,
  title={Turing Instability and Spatiotemporal Pattern Formation Induced by Nonlinear Reaction Cross-Diffusion in a Predator--Prey System with Allee Effect},
  author={Shao, Yangyang and Meng, Yan and Xu, Xinyue},
  journal={Mathematics},
  volume={10},
  number={9},
  pages={1500},
  year={2022},
  publisher={MDPI}
}

@article{van2021pattern,
  title={Pattern formation from spatially heterogeneous reaction--diffusion systems},
  author={Van Gorder, Robert A},
  journal={Philosophical Transactions of the Royal Society A},
  volume={379},
  number={2213},
  pages={20210001},
  year={2021},
  publisher={The Royal Society}
}

@article{parker1991eddy,
  title={Eddy diffusion of phytoplankton and nutrients: estimating coefficients from simulated and observed vertical distributions},
  author={Parker, RA},
  journal={Journal of Plankton Research},
  volume={13},
  number={4},
  pages={815--830},
  year={1991},
  publisher={Oxford University Press}
}

@article{villar2025designing,
  title={Designing reaction-cross-diffusion systems with {T}uring and wave instabilities},
  author={Villar-Sep{\'u}lveda, Edgardo and Champneys, Alan R and Krause, Andrew L},
  journal={Journal of Mathematical Biology},
  volume={91},
  number={4},
  pages={37},
  year={2025},
  publisher={Springer}
}

@article{villar2023general,
  title={General conditions for {T}uring and wave instabilities in reaction-diffusion systems},
  author={Villar-Sep{\'u}lveda, Edgardo and Champneys, Alan R},
  journal={Journal of Mathematical Biology},
  volume={86},
  number={3},
  pages={39},
  year={2023},
  publisher={Springer}
}

@article{mondal2023impact,
  title={Impact of critical eddy diffusivity on seasonal bloom dynamics of Phytoplankton in a global set of aquatic environments},
  author={Mondal, Arpita and Banerjee, Sandip},
  journal={Scientific Reports},
  volume={13},
  number={1},
  pages={17141},
  year={2023},
  publisher={Nature Publishing Group UK London}
}

@article{nisbet1976model,
  title={Model of material cycling in a closed ecosystem},
  author={Nisbet, RM and Gurney, WSC},
  journal={Nature},
  volume={264},
  number={5587},
  pages={633--634},
  year={1976},
  publisher={Nature Publishing Group UK London}
}

@article{ulanowicz1972mass,
  title={Mass and energy flow in closed ecosystems},
  author={Ulanowicz, Robert E},
  journal={Journal of Theoretical Biology},
  volume={34},
  number={2},
  pages={239--253},
  year={1972},
  publisher={Elsevier}
}

@article{satnoianu2000turing,
  title={Turing instabilities in general systems},
  author={Satnoianu, Razvan A and Menzinger, Michael and Maini, Philip K},
  journal={Journal of Mathematical Biology},
  volume={41},
  number={6},
  pages={493--512},
  year={2000},
  publisher={Springer}
}

@article{giricheva2019spatiotemporal,
  title={Spatiotemporal dynamics of an NPZ model with prey-taxis and intratrophic predation},
  author={Giricheva, Evgeniya},
  journal={Nonlinear Dynamics},
  volume={95},
  number={2},
  pages={875--892},
  year={2019},
  publisher={Springer}
}

@inproceedings{Auto,
author="E.J. Doedel",
title = "AUTO: A program for the automatic bifurcation analysis of
autonomous systems",
booktitle = "Proc. 10th Manitoba Conf. on Num. Math. and Comp.,
Univ. of Manitoba, Winnipeg, Canada",
pages = "265-284",
year=1981 }

@book{XPPAUT,
author="G.B. Ermentrout",
title="Similating, Analyzing and Animating Dynamical Systems:
A Guide to XPPAUT for Researcher and Students.",
publisher="SIAM",
address="{Philadelphia, PA}",
year=2002}

\appendix
\section{Existence and Stability of Equilibrium Solutions of the ODE model}\label{app:eqm}
Here we consider conditions to guarantee the existence and stability of biologically feasible equilibria. Many of the results presented below follow from  \cite{kloosterman2014closed}, since eq.~\eqref{eq:NPZD_ODE_system} corresponds to taking a particular distribution, the weak gamma distribution, in the model considered in \cite{kloosterman2014closed}. We repeat some details here for completeness and to put the results into the NPZD notation.

\subsection{Existence and Uniqueness}
Note that the equilibrium equations \eqref{eq:eqmdef1}-\eqref{eq:eqmdef3} are satisfied by setting $P^*=Z^*=D^*=0$. It follows from the conservation law~\eqref{eq:eqmdef4} that a unique trivial equilibrium $(N_T,0,0,0)$ exists for each $N_T$ and is biologically feasible if $N_T>0.$

Now consider equilibria of the form $E_{NPD}=[N^*,P^*,0,D^*]$, where $P^*,D^*$ are nonzero. Setting $Z^*=0$ satisfies eq.~\eqref{eq:eqmdef3}. Solving eqs.~\eqref{eq:eqmdef1}-\eqref{eq:eqmdef2} gives
 \begin{equation} N^*=\frac{\lambda}{\nu-\lambda},\quad D^*=\frac{\lambda}{\alpha}P^*.
 \label{eq:NP_P}\end{equation}
 Substituting this into the conservation law \eqref{eq:eqmdef4} gives
 \begin{equation} \frac{\lambda}{\nu-\lambda}+ \left(1+\frac{\lambda}{\alpha}\right)P^*=N_T\label{eq:NPcons}
 \end{equation}
It follows that the components of the equilibrium are given by 
\[
N^*=\frac{\lambda}{\nu-\lambda}, \quad
P^*=\frac{N_T-\frac{\lambda}{\nu-\lambda}}{1+\frac{\lambda}{\alpha}},\quad
D^*=\frac{N_T-\frac{\lambda}{\nu-\lambda}}{1+\frac{\alpha}{\lambda}}.
\]
Clearly, for each value of $N_T$ there is a unique equilibrium.
In order for this equilibrium to be biologically feasible, $N^*>0$ requires $\nu>\lambda$ while $P^*,D^*>0$ require
\[N_T>\frac{\lambda}{\nu-\lambda}\stackrel{def}{=}N_{T1}.
\]
Biologically the first condition means the maximum growth rate of the phytoplankton is larger that the maximum death rate, while the second condition is means that the total biomass in the system sufficiently large. Note that the value of $N^*$ is independent of $N_T$ while $P^*$ and $D^*$ are linear, increasing functions of $N_T$ 
 

 Now consider equilibria of the form $[N^*,P^*,Z^*,D^*]$. From equations \eqref{eq:eqmdef1}-\eqref{eq:eqmdef3} we have
 \begin{equation} 
 P^*=\frac{\delta}{g-\delta},\quad D^*=
 \frac{\nu}{\alpha}P^*\frac{N^*}{N^*+1},\quad Z^*=\frac{P^*}{\delta}\left(\nu \frac{N^*}{N^*+1}-\lambda\right).
 \label{eq:ZDofN}
 \end{equation}
 Thus we see that a necessary condition for the biological feasibility of this equilibrium is  $g>\delta$.
 Substituting the expressions above into \eqref{eq:eqmdef4} gives 
 \begin{equation}
 N^*+P^*+\frac{\nu}{\alpha}P^*\frac{N^*}{N^*+1}+\frac{P^*}{\delta}\left(\nu \frac{N^*}{N^*+1}-\lambda\right)=N_T, \label{eq:NPZ_cons}
 \end{equation}
 which can be rearranged into a quadratic for $N^*$:
 \begin{equation} N^{*2}+\left(1+\left(1-\frac{\lambda}{\delta}+\frac{\nu}{\delta}+\frac{\nu}{\alpha}\right)P^*-N_T\right)N^*+\left(1-\frac{\lambda}{\delta}\right)P^*-N_T=0.\label{eq:Nquad}
 \end{equation}
 From this it follows that there is a unique positive solution for $N^*$ if and only if 
 \begin{equation} N_T>\left(1-\frac{\lambda}{\delta}\right)P^*=\frac{\delta-\lambda}{g-\delta},\label{eq:tcond}\end{equation}
 which is gives a second necessary condition for biological feasibility. The final condition comes from $Z^*>0$ which is true if and only if
 \begin{equation} \frac{N^*}{N^*+1}>\frac{\lambda}{\nu}\quad\Leftrightarrow\quad N^*>\frac{\lambda}{\nu-\lambda}=N_{T1}.\label{eq:tcond2}\end{equation}
 Substituting $N=N^*_{NP}$ into \eqref{eq:NPZ_cons} gives
 \begin{equation} N_T=N_{T1}+\left(1+\frac{\lambda}{\alpha}\right)P^*=\frac{\lambda}{\nu-\lambda}+\left(1+\frac{\lambda}{\alpha}\right)\frac{\delta}{g-\delta}\stackrel{def}{=}N_{T2}.\label{eq:NT2app}
 \end{equation}
 Note that $N_T>N_{T2}$ implies \eqref{eq:tcond}.

 Rewriting \eqref{eq:Nquad} as
$N^*+\beta N^*+\gamma=0$,
the positive root of eq.~\eqref{eq:Nquad} can be written
 \[ N^*=\frac{-\beta+\sqrt{\beta^2-4\gamma}}{2}.\]
Differentiating eq.~\eqref{eq:Nquad} and rearranging shows that
\[ \frac{dN^*}{dN_T}=\frac{N^*+1}{2N^*+\beta}=\frac{N^*+1}{\sqrt{\beta^2-4\gamma}}>0.\]
Thus $N^*$ is an increasing function of $N_T$. This means that $N_T>N_{T2}$ implies \eqref{eq:tcond2}. Further, since $f(N^*)=\frac{N^*}{N^*+1}$ is an increasing function of $N^*$, it follows from eq.~\eqref{eq:ZDofN} that $Z^*$ and $D^*$ are increasing functions of $N_T$. 

 In summary, if $g>\delta$, there is a unique biologically feasible equilibrium $(N^*,P^*,Z^*,D^*)$ for each $N_T$ satisfying $N_T>N_{T2}.$ $P^*$ is independent of $N_T$, while $N^*,Z^*,D^*$ are increasing functions of $N_T$.
 \subsection{Stability}

 The Jacobian of the ODE system \eqref{eq:NPZD_ODE_system} at the equilibrium $[N_T, 0, 0, 0]$ is given by:
\begin{equation}
    \label{eq:npzd_jacobian}
    J_{N}  
    = \begin{bmatrix}
        0 & -\nu\frac{N_T}{N_T+1} & 0 & \alpha \\
        0 & \nu\frac{N_T}{N_T+1}-\lambda & 0 & 0 \\
        0 & 0 & -\delta & 0 \\
        0 & \lambda & \delta & -\alpha
    \end{bmatrix}.
\end{equation}
It is easy to show that the eigenvalues of $J_N$ are $0,-\delta,-\alpha, \nu\frac{N_T}{N_T+1}-\lambda$.  Thus this equilibrium will be stable if $0<N_T<N_{T1}$  and unstable if $N_T>N_{T1}$. It is nonhyperbolic when$N_T=N_{T1}$ due the presence of a second zero eigenvalue. 

 The Jacobian of the ODE system \eqref{eq:NPZD_ODE_system} at the equilibrium $[N^*, P^*, 0, D^*]$ is given by:
\begin{equation}
    \label{eq:npzd_jacobian}
    J_{NP}  
    = \begin{bmatrix}
        -A & -\lambda & 0 & \alpha \\
        A & 0 & -g\frac{P^*}{P^*+1} & 0 \\
        0 & 0 & g\frac{P^*}{P^*+1}-\delta & 0 \\
        0 & \lambda & \delta & -\alpha
    \end{bmatrix},
\end{equation}
where $A=\nu P^*(1+N^*)^{-2}.$
The eigenvalues of $J_{NP}$ are $0,\mu_{\pm},g\frac{P^*}{P^*+1}-\delta$ where
\[ \mu_{\pm}=\frac{1}{2}\left(-(A+\alpha)\pm \sqrt{(A+\alpha)^2-4A(\alpha+\lambda)}\right).\] Note that $\alpha,\lambda$ are positive parameters, and $A>0$ when the equilibrium is biologically feasible. It follows that $\Re(\mu_{\pm})<0$. Further, it follows from eqs.~\eqref{eq:NPcons} and ~\eqref{eq:NT2app} that
\[ g\frac{P^*}{P^*+1}>\delta \quad \Leftrightarrow N_T>N_{T2}.\]
Thus this equilibrium will be stable if $N_{T1}<N_T<N_{T2}$ and unstable if $N_T>N_{T2}$. 
Note that this is stronger than the result obtained in \cite[Proposition 4]{kloosterman2014closed} for a general distribution of delays. This equilibrium is nonhyperbolic when $N_T=N_{T2}$ due the presence of a second zero eigenvalue.

\end{document}